\documentclass[a4paper,11pt]{article}

\pdfoutput=1 

\usepackage{jheppub} 

\usepackage{slashed}
\usepackage{subcaption}
\usepackage{xcolor}
\usepackage{cleveref}
\usepackage[T1]{fontenc} 
\usepackage[utf8]{inputenc}

\newcommand{\cA}{\mathcal{A}}
\newcommand{\cM}{\mathcal{M}}

\title{\boldmath On-shell heavy particle effective theories}

\author[a]{Rafael Aoude,}
\author[b]{Kays Haddad,}
\author[b]{and Andreas Helset}

\affiliation[a]{PRISMA$^{+}$ Cluster of Excellence \& Institute of Physics, \\
Johannes Gutenberg-Universit\"{a}t Mainz, 55099 Mainz, Germany}
\affiliation[b]{Niels Bohr International Academy and Discovery Center,
Niels Bohr Institute, University of Copenhagen, Blegdamsvej 17,
DK-2100 Copenhagen, Denmark}

\emailAdd{aoude@uni-mainz.de}
\emailAdd{kays.haddad@nbi.ku.dk}
\emailAdd{ahelset@nbi.ku.dk}

\abstract{We introduce on-shell variables for Heavy Particle Effective Theories (HPETs) with the goal of extending Heavy Black Hole Effective Theory to higher spins and of facilitating its application to higher post-Minkowskian orders. These variables inherit the separation of spinless and spin-inclusive effects from the HPET fields, resulting in an explicit spin-multipole expansion of the three-point amplitude for any spin.
By matching amplitudes expressed using the on-shell HPET variables to those derived from the one-particle effective action, we find that the spin-multipole expansion of a heavy spin-$s$ particle corresponds exactly to the multipole expansion (up to order $2s$) of a Kerr black hole, that is, without needing to take the infinite spin limit.
Finally, we show that tree-level radiative processes with same-helicity bosons emitted from a heavy spin-$s$ particle exhibit a spin-multipole universality.
}
\subheader{\normalsize
        \vspace*{-3.7em}
        \begin{flushright}
                SAGEX-20-02-E \\
                MITP/20-002
        \end{flushright}
}

\allowdisplaybreaks

\begin{document} 
\maketitle
\flushbottom

\section{Introduction}\label{sec:Intro}

The relationship between quantum scattering amplitudes and classical physics has enjoyed a surge of attention in recent years, in large part due to the observation of gravitational waves by the LIGO and Virgo collaborations as of 2015 \cite{LIGOGW}. Motivating studies in this direction has been the realization that perturbative techniques from quantum field theory are well suited to the computation of the complementary post-Newtonian (PN) and post-Minkowskian (PM) expansions of the binary inspiral problem in General Relativity (GR). 
Indeed, the effective field theory (EFT) of GR \cite{Donoghue:1993eb,Donoghue:1994dn} has been used extensively to compute classical corrections 
to the 
gravitational potential \cite{Donoghue:1994dn,BjerrumBohr:2002kt,Holstein:2008sx,Bjerrum-Bohr:2013bxa,Neill:2013wsa,Vaidya:2014kza,Cachazo:2017jef,Guevara:2017csg,Chung:2018kqs}. Furthermore, effective-field-theoretic methods have been used to develop EFTs for gravitationally interacting objects whose operator expansions are tailored to computing terms in the PN approximation \cite{Goldberger:2004jt,Porto:2005ac,Levi:2015msa,Levi:2015ixa,Levi:2018nxp,Levi:2019kgk}. 
In fact, the current state-of-the-art computations involving spin from the PN approach were
performed in refs.~\cite{Levi:2015ixa,Levi:2019kgk} using the EFT of ref.~\cite{Levi:2015msa}.
On the PM side, it has also recently been shown that quantum scattering amplitudes can be used to extract fully relativistic information about the classical scattering process \cite{Bjerrum-Bohr:2018xdl,Cheung:2018wkq,Guevara:2018wpp,Zvi3PM,Bern:2019crd,Cristofoli:2019neg,Cristofoli:2019ewu,Kosower:2018adc,Maybee:2019jus,Guevara:2019fsj}.
Moreover, a direct relationship between the scattering amplitude and the scattering angle has been uncovered in refs.~\cite{Damour:2016gwp,Damour:2017zjx,Kalin:2019rwq,Bjerrum-Bohr:2019kec}.\footnote{We thank Andrea Cristofoli for bringing earlier work on this relationship to our attention.}
All of these developments suggest that the $2\rightarrow2$ gravitational scattering amplitude encodes information that is crucial for the understanding of classical gravitational binary systems, to all loop orders \cite{Donoghue:1993eb,Holstein:2004dn}.

Various methods exist for identifying the classical component of a scattering amplitude \cite{Cachazo:2017jef,Guevara:2017csg,Kosower:2018adc,Maybee:2019jus}. Towards this same end, Heavy Black Hole Effective Theory (HBET) was recently formulated by Damgaard and two of the present authors in ref.~\cite{Damgaard:2019lfh} with the aim of streamlining the extraction of classical terms from gravitational scattering amplitudes. It was shown there that the operator expansion of HBET is equivalent to an expansion in $\hbar$. Exploiting this fact, the authors were able to identify which HBET operators can induce classical effects at arbitrary loop order, and the classical portion of the $2\rightarrow2$ amplitude was computed up to one-loop order for spins $s\leq1/2$.
These results were obtained using Lagrangians and Feynman diagram techniques which, while tractable at the perturbative orders and spins considered, become non-trivial and computationally unwieldy to extend to higher spins or loop orders. Nevertheless, the separation of classical and quantum effects and the observed separation of spinless and spin-inclusive effects are desirable features of the EFT that will prove quite convenient when cast as part of a more user-friendly formalism.

We aim in this paper to present such a formalism that will allow the extension of HBET to higher spins
and to facilitate its application to higher loop orders.
A means to do so comes in the formalism presented in ref.~\cite{Arkani-Hamed:2017jhn}.
Spinor-helicity variables were presented there that describe the scattering
of massive matter with arbitrary spin.
Based solely on kinematic considerations, these variables were used to construct
the most general three-point amplitude for a massive spin-$s$ particle
emitting a massless boson with a given helicity.
In this most general amplitude, the term that is best behaved in the UV limit is termed the minimal coupling amplitude. When $s\leq1/2$ it reduces
to the three-point amplitude arising from the relevant Lagrangian
that is minimally coupled in the sense of covariantized derivatives.
This terminology is preserved for higher spins;
the minimal coupling amplitude for a general spin-$s$ particle is a tensor product
of $2s$ factors of spin-1/2 minimal coupling amplitudes.
Note that this definition of minimal coupling generally differs from the typical
definition from the Lagrangian perspective.
Phenomenologically, these minimal coupling amplitudes are those that
produce a gyromagnetic ratio of $g=2$ for all spins \cite{Chung:2018kqs,Holstein:2006wi,NaturalgFactor}.

This minimal coupling amplitude has proven to be quite useful in the study of
classical Kerr black holes, which have been shown to couple minimally to gravity
\cite{Guevara:2017csg,Guevara:2018wpp,Chung:2018kqs,Guevara:2019fsj,Arkani-Hamed:2019ymq,Chung:2019duq}.
Such a description of Kerr black holes is in fact not immediately exact when using the variables of ref.~\cite{Arkani-Hamed:2017jhn}
due to the difference between the momenta of the initial and final states, 
leading to an ill-defined matrix element of the spin-operator.
This gap has been overcome using various methods in the above references. 
However we will show that expressing the degrees of freedom of HBET in on-shell 
variables reduces the discrepancy to a mere choice of the kinematics. 
The appropriate kinematics can sometimes be imposed (when a process is described by diagrams with no internal matter lines), but are always recovered in the classical limit; $\hbar\rightarrow0$.

In this paper, we express the asymptotic states of Heavy Particle Effective Theories (HPETs) --- the collection of effective field theories treating large mass particles --- using the massive on-shell spinor-helicity variables of ref.~\cite{Arkani-Hamed:2017jhn}. 
An explicit $\hbar$ expansion will arise from these variables, which makes simple the task of taking classical limits of amplitudes.
Such an expression of the asymptotic states of HPET will also lead to an explicit separation of spinning and spinless effects in the three-point minimal coupling amplitude.
From the lens of the classical gravitational scattering of two spinning black holes, this results in the finding that the asymptotic states of HPET are naturally identified with a Kerr black hole with truncated spin-multipole expansion.

Our construction will also allow us to gain insight into this class of effective field theories. We will derive a conjecture for the three-point amplitude arising from an arbitrary HPET, and posit a form for this same amplitude for heavy matter of any spin. Then, in the appendices, we comment on the link between reparameterization invariance of a momentum and its little group, and finally compute the operator projecting onto a heavy particle of spin $s\leq2$, the derivation of which can be extended to general spin. 

The layout of this paper is as follows. We begin with a very brief review of HPETs in Section~\ref{sec:TopDown}. Also, we introduce on-shell variables that describe the heavy field. The three-point amplitudes of HPETs are analyzed in Section~\ref{sec:3Pt}. In particular, we construct the three-point amplitude of HPET resummed to all orders in the expansion parameter. Furthermore, the construction of ref.~\cite{Arkani-Hamed:2017jhn} provides a method of extending HPET amplitudes to arbitrary spin.
In Section~\ref{sec:OnePartEffAct}, we interpret the on-shell HPET variables as Kerr black holes with truncated spin-multipole expansions, and show that heavy spin-$s$ particles possess the same spin-multipole expansion as a Kerr black hole, up to the $2s^{\rm th}$ multipole. This is in contrast to previous work \cite{Chung:2018kqs,Chung:2019duq}, which found that minimally coupled particles possess the same spin multipoles as Kerr black holes only in the infinite spin limit.
Section~\ref{sec:OnShellAmps} is dedicated to the computation of on-shell amplitudes, and we show the simplicity of taking the classical limit of an amplitude when it is expressed in on-shell HPET variables. 
The main body of the paper is concluded in Section~\ref{sec:Conclusion}. 
Our conventions are summarized in Appendix~\ref{sec:Conventions}. 
The question of the uniqueness of the constructed variables is addressed in Appendix~\ref{sec:Uniqueness}.
We then relate the little group of a momentum $p$ to its invariance under the HPET reparameterization (see Section~\ref{sec:TopDown}) in \Cref{sec:RepLG}.
In Appendix~\ref{sec:Propagators} we use spin-$s$ polarization tensors for heavy particles to explicitly construct propagators and projection operators for heavy particles with spins $s\leq2$. 
We then use these results to conjecture the forms of the projection operators for arbitrary spin. 
Finally, we describe in Appendix~\ref{sec:HPETMatch} the forms of the spin-1/2 HPET Lagrangians that must be used to match to the on-shell minimal coupling amplitudes. 
We also show there that the three-point amplitude derived from a Lagrangian for a heavy spin-1 particle is reproduced by the extension of the variables to arbitrary spin in \Cref{sec:3Pt}.

\section{Effective theories with heavy particles}\label{sec:TopDown}

When describing a scattering process in which the transfer momentum, $q^{\mu}$, is small compared to the mass of one of the scattered particles, $m$, 
we can exploit the separation of scales by expanding in the small parameter $|q|/m$. Heavy Quark Effective Theory (HQET) \cite{Georgi:1990um,Bodwin:1994jh,HQETRev} is the effective field theory that employs this expansion in the context of QCD, with HBET being its gravitational analog.
Central to the separation of scales is 
the decomposition of the momentum of the heavy particle as
\begin{align}
    p^{\mu}&=mv^{\mu}+k^{\mu},\label{eq:HQETMomDecomp}
\end{align}
where $v^{\mu}$ is the (approximately constant) four-velocity ($v^{2}=1$) of the heavy particle, 
and $k^{\mu}$ is a residual momentum that parameterizes the energy of the interaction; 
it is therefore comparable in magnitude to the momentum transfer, $|k^{\mu}|\sim |q^{\mu}|$. 
When decomposed in this way, the on-shell condition, $p^{2}=m^{2}$, is equivalent to 
\begin{align}\label{eq:HPETOnShell}
    v\cdot k&=-\frac{k^{2}}{2m}.
\end{align}
%
As was argued in ref.~\cite{Damgaard:2019lfh}, using results from ref.~\cite{Kosower:2018adc}, the residual momentum scales with $\hbar$ in the limit $\hbar\rightarrow0$. We discuss the counting of $\hbar$ in \Cref{sec:hbarCount}.

With some background about the construction and motivation behind HPETs, we introduce in this section on-shell variables that describe spin-1/2 HPET states. Then, the transformation of these variables under a reparameterization of the momentum \cref{eq:HQETMomDecomp} is given. We end the section by defining the spin operator for heavy particles.


\subsection{On-shell HPET variables}

%
The spinors $u^{I}_{v}(p)$ that describe the particle states of HPET are related to the Dirac spinors $u^{I}(p)$ via \cite{Damgaard:2019lfh}
\begin{align}
    u^{I}_{v}(p)&=
    \left(\frac{\mathbb{I}+\slashed{v}}{2}\right)u^{I}(p) = 
    \left(\mathbb{I}-\frac{\slashed{k}}{2m}\right)u^{I}(p),\label{eq:HQETSpinor}
\end{align}
where $I$ is an $SU(2)$ little group index, and $v^{\mu}$ and $k^{\mu}$ are defined in \cref{eq:HQETMomDecomp}. 
The operator $P_{+}\equiv\frac{1+\slashed{v}}{2}$ is the projection operator
that projects on to the heavy particle states.
Writing the Dirac spinor in terms of massive on-shell spinors $|\mathbf{p}\rangle_{\alpha}$ and $|\mathbf{p}]^{\dot\alpha }$, 
we define on-shell variables for the HPET spinor field:
\begin{align}
    \begin{pmatrix}
	    |\mathbf{p}_{v}\rangle \\
    |\mathbf{p}_{v}]
    \end{pmatrix}&=\left(\mathbb{I}-\frac{\slashed{k}}{2m}\right)\begin{pmatrix}
	    |\mathbf{p}\rangle \\
    |\mathbf{p}]
    \end{pmatrix}.\label{eq:HQETOnShellVars}
\end{align}
The bold notation for the massive on-shell spinors was introduced in ref.~\cite{Arkani-Hamed:2017jhn}, and represents symmetrization over the little group indices.
We refer to the on-shell variables of ref.~\cite{Arkani-Hamed:2017jhn} as the traditional on-shell variables, and those introduced here as the on-shell HPET variables. 
The on-shell HPET variables are labelled by their four-velocity $v$.
We emphasize that the relation between the traditional and HPET on-shell variables is exact in $k/m$.
See \Cref{sec:Conventions} for conventions.

When working with heavy particles, the Dirac equation is replaced by the relation $\slashed{v}u^I_{v} = u^I_{v}$, which can be seen by multiplying the first equation in \cref{eq:HQETSpinor} by $\slashed{v}$. This 
relates the on-shell HPET variables in different bases through
%
%
\begin{subequations}\label{eq:HQETMix}
\begin{align}
v_{\alpha\dot{\beta}}|\mathbf{p}_{v}]^{\dot{\beta}}=|\mathbf{p}_{v}\rangle_{\alpha},
&\quad v^{\dot{\alpha}\beta}|\mathbf{p}_{v}\rangle_{\beta}=|\mathbf{p}_{v}]^{\dot{\alpha}}, \\
[\mathbf{p}_{v}|_{\dot{\alpha}}v^{\dot{\alpha}\beta}=-\langle \mathbf{p}_{v}|^{\beta },
	&\quad \langle \mathbf{p}_{v}|^{\alpha }v_{\alpha\dot{\beta}}=-[\mathbf{p}_{v}|_{\dot{\beta}}.
\end{align}
\end{subequations}
%
%
%
%
We associate the momentum $p_{v}^\mu$ with the on-shell HPET spinors, where
\begin{align}\label{eq:pHMom}
    \slashed{p}_{v}&=\begin{pmatrix}
    0 & |p_{v}\rangle^{I}{}_{I}[p_{v}| \\
    |p_{v}]_{I}{}^{I}\langle p_{v}| & 0
    \end{pmatrix}
    =m_{k}\slashed{v},
\end{align}
and
\begin{align}\label{eq:HQETVarMass}
    m_{k}&\equiv\left(1-\frac{k^{2}}{4m^{2}}\right)m.
\end{align}
We see that the momentum $p_v^\mu$ is proportional to $v^{\mu}$, regardless of the residual momentum. 
The momentum $p_{v}^{\mu}$ is related to the momentum $p^{\mu}$ through
\begin{align}
    P_+\slashed{p}_v &= P_+ \slashed{p}P_+.
\end{align}
The on-shell HPET variables naturally describe heavy particles in a context with no anti-particles. To see this, note that the relation between the HPET spinor and the Dirac spinor in \cref{eq:HQETOnShellVars} can be inverted
\cite{Finkemeier:1997re}
%
\begin{align}
	\label{eq:spinorInverted}
    u^{I}(p)&=\left(\mathbb{I}-\frac{\slashed{k}}{{2m}}\right)^{-1}u^{I}_{v}(p)\notag \\
    &=\left[1+\frac{1}{2m}\left(1+\frac{k\cdot v}{2m}\right)^{-1}(\slashed{k}-k\cdot v)\right]u^{I}_{v}(p).
\end{align}
%
In the free theory, this corresponds to the relation between the fields in the full and effective theories 
once the heavy anti-field 
has been integrated out by means of its equation of motion. Thus, \cref{eq:HQETSpinor} is equivalent to integrating out heavy anti-particle states.

\subsection{Reparameterization}

There is an ambiguity in the choice of $v$ and $k$ in the decompositon of the momentum in \cref{eq:HQETMomDecomp}.
The momentum is invariant under reparameterizations of $v$ and $k$ of the forms
\begin{align}
    (v,k)&\rightarrow(w,k^{\prime})\equiv\left(v+\frac{\delta k}{m},k-\delta k\right),\label{eq:Reparam}
\end{align}
%
%
where $|\delta k|/m\ll1$ and $(v+\delta k/m)^{2}=1$.
Given that observables can only depend on the total momentum,
observables computed in heavy particle effective theories must be invariant under this 
reparameterization \cite{Luke:1992cs,CHEN1993421,Finkemeier:1997re}.
In particular, the $S$-matrix is reparameterization invariant.

The on-shell HPET variables transform under the reparameterization of the momentum in \cref{eq:Reparam}.
The HPET spinors $u^I_{v}(p)$ and $u^I_{w}(p)$ are related through
\begin{align}
    u^I_{v}(p)&=\frac{1+\slashed{v}}{2}u^I(p)\notag \\
    &=\frac{1+\slashed{v}}{2}\left[1+\frac{1}{2m}\left(1+\frac{k^\prime\cdot w}{2m}\right)^{-1}(\slashed{k}^\prime-k^\prime\cdot w)\right]u^I_{w}(p),
\end{align}
where the second line is simply \cref{eq:spinorInverted} with $(v,k)\rightarrow(w,k^\prime)$. 
Rewriting this in terms of the on-shell HPET variables, we find
\begin{subequations}\label{eq:ReparamTrans}
\begin{align}
|\mathbf{p}_{v}\rangle&=\left(1-\frac{k^{\prime2}}{4m^{2}}\right)^{-1}\left[\left(1-\frac{k^{2}}{4m^{2}}+\frac{\slashed{k}\delta\slashed{k}}{4m^{2}}\right)|\mathbf{p}_{w}\rangle-\frac{\delta\slashed{k}}{2m}|\mathbf{p}_{w}]\right], \\
    |\mathbf{p}_{v}]&=\left(1-\frac{k^{\prime2}}{4m^{2}}\right)^{-1}\left[\left(1-\frac{k^{2}}{4m^{2}}+\frac{\slashed{k}\delta\slashed{k}}{4m^{2}}\right)|\mathbf{p}_{w}]-\frac{\delta\slashed{k}}{2m}|\mathbf{p}_{w}\rangle\right].
\end{align}
Similarly,
\begin{align}
	\langle \mathbf{p}_{v}|&=\left(1-\frac{k^{\prime2}}{4m^{2}}\right)^{-1}
	\left[\langle \mathbf{p}_{w}|\left(1-\frac{k^{2}}{4m^{2}}+\frac{\delta\slashed{k}\slashed{k}}{4m^{2}}\right)+[\mathbf{p}_{w}|\frac{\delta\slashed{k}}{2m}\right], \\
		[\mathbf{p}_{v}|&=\left(1-\frac{k^{\prime2}}{4m^{2}}\right)^{-1}
			\left[[\mathbf{p}_{w}|\left(1-\frac{k^{2}}{4m^{2}}+\frac{\delta\slashed{k}\slashed{k}}{4m^{2}}\right)+\langle \mathbf{p}_{w}|\frac{\delta\slashed{k}}{2m}\right].
\end{align}
\end{subequations}
The transformed spinors $|\mathbf{p}_w\rangle$ and $|\mathbf{p}_w]$ are related to the traditional
on-shell variables via \cref{eq:HQETOnShellVars}, with the replacement $k\rightarrow k^\prime$.

This transformation is singular at the point where the new residual momentum has magnitude squared $k^{\prime2}=4m^{2}$. This pole is ubiquitous when using these variables, and signals the point where fluctuations of the matter field are energetic enough to allow for pair-creation. As we have integrated out the anti-particle through \cref{eq:HQETSpinor}, such energies are outside the region of validity of this formalism. In fact, the working assumption of the formalism is that the residual momentum is small compared to the mass, so one would expect the formalism to lose predictive power well before this point.


\subsection{Spin operator}

We identify the spin operator with the Pauli-Lubanski pseudovector,
\begin{align}
	\label{eq:PauliLub}
    S^{\mu}=-\frac{1}{2m}\epsilon^{\mu\nu\alpha\beta}p_{\nu}J_{\alpha\beta},
\end{align}
where $J^{\mu\nu}$ is the generator of rotations, $p^{\mu}$ is the momentum with respect to which the operator is defined, and $m^{2}=p^{2}$. 
For our purposes, it will be convenient to choose $p^{\mu}=p_{v}^{\mu}$: this ensures that, irrespective of the value of the residual momentum, the momentum $p_{v}^{\mu}=m_{k}v^{\mu}$ will always be orthogonal to the spin operator. Thus, $S^{\mu}$ is the spin vector of a particle with velocity $v^{\mu}$ and any value of residual momentum. With this choice for the reference momentum, the spin-operator is
\begin{align}
    S^{\mu}=-\frac{1}{2}\epsilon^{\mu\nu\alpha\beta}v_{\nu}J_{\alpha\beta}.
\end{align}
%
Its action on irreducible representations of $SL(2,\mathbb{C})$ is \cite{Chung:2018kqs}
\begin{subequations}
\begin{align}
    {\left(S^{\mu}\right)_{\alpha}}^{\beta}&=\frac{1}{4}\left[(\sigma^{\mu})_{\alpha\dot{\alpha}}v^{\dot{\alpha}\beta}-v_{\alpha\dot{\alpha}}(\bar{\sigma}^{\mu})^{\dot{\alpha}\beta}\right]\label{eq:ChiralSpin}, \\
    {\left(S^{\mu}\right)^{\dot\alpha}}_{\dot\beta}&=-\frac{1}{4}\left[(\bar{\sigma}^{\mu})^{\dot{\alpha}\alpha}v_{\alpha\dot\beta}-v^{\dot\alpha\alpha}(\sigma^{\mu})_{\alpha\dot\beta}\right]\label{eq:AntiChiralSpin}.
\end{align}
\end{subequations}
These two representations of the spin-vector are related via
\begin{align}
  {\left(S^{\mu}\right)_{\alpha}}^{\beta}&=v_{\alpha\dot{\alpha}}{\left(S^{\mu}\right)^{\dot\alpha}}_{\dot\beta}v^{\dot{\beta}\beta},\quad {\left(S^{\mu}\right)^{\dot\alpha}}_{\dot\beta}=v^{\dot{\alpha}\alpha}{\left(S^{\mu}\right)_{\alpha}}^{\beta}v_{\beta\dot{\beta}}.
\end{align}
On three-particle kinematics, the spin-vector can be written more compactly by introducing the $x$ factor for a massless momentum $q$ \cite{Arkani-Hamed:2017jhn},
\begin{subequations}
\begin{align}
    mx\langle q|&\equiv[q|p_{1}, \\
    \Rightarrow mx^{-1}[q|&=\langle q|p_{1}.
\end{align}
\end{subequations}
Using this, 
when the initial residual momentum is $k=0$, we can re-express the contraction $q\cdot S$ as
\begin{subequations}\label{eq:SpinZeroRes}
\begin{align}
    {\left(q\cdot S\right)_{\alpha}}^{\beta}&=\frac{x}{2}|q\rangle\langle q|, \\
    {\left(q\cdot S\right)^{\dot\alpha}}_{\dot\beta}&=-\frac{x^{-1}}{2}|q][q|.
\end{align}
\end{subequations}
For general initial residual momentum, we find an additional term:
\begin{subequations}\label{eq:SpinGenRes}
\begin{align}
    {\left(q\cdot S\right)_{\alpha}}^{\beta}&=\frac{1}{4}\left(2x|q\rangle\langle q|+\frac{1}{m}{[k,q]_{\alpha}}^{\beta}\right), \\
    {\left(q\cdot S\right)^{\dot\alpha}}_{\dot\beta}&=-\frac{1}{4}\left(2x^{-1}|q][q|+\frac{1}{m}{[k,q]^{\dot\alpha}}_{\dot\beta}\right).
\end{align}
\end{subequations}
Note that \cref{eq:SpinGenRes} reduces to \cref{eq:SpinZeroRes} when $k=0$.

When choosing the reference momentum to be $p_{v}^{\mu}$, we can identify the
spin-vector with the classical spin-vector of a Kerr black-hole with classical
momentum $p_{\text{Kerr}}^{\mu}=\frac{m}{m_{k}}p_{v}^{\mu}$. This is because the
Lorentz generator in \cref{eq:PauliLub} can be replaced with the black hole
spin-tensor $S^{\mu\nu}=J_{\perp}^{\mu\nu}$ which satisfies the condition 
\cite{tulczyjew1959motion,Levi:2015msa}
\begin{align}
    p_{\text{Kerr}}^{\mu}S_{\mu\nu}=0,
\end{align}
known as the spin supplementary condition.

In ref.~\cite{Damgaard:2019lfh}, the spin vector was defined as
\begin{align}
	S^{\mu}_{v}\equiv\frac{1}{2}\bar{u}_{v}(p_2)\gamma_{5}\gamma^{\mu}u_{v}(p_1),
\end{align}
and it was found that this spin vector satisfied the relation
\begin{align}
	\bar{u}_{v}(p_2)\sigma^{\mu\nu}u_{v}(p_1)=-2\epsilon^{\mu\nu\alpha\beta}v_{\alpha}S_{v\beta}.
\end{align}
We can therefore relate these two definitions of the spin vector:
\begin{align}\label{eq:SpinDefRel}
	S^{\mu}_{v}&=\bar{u}_{v}(p_2)S^{\mu}u_{v}(p_1)=-2\langle\mathbf{2}_{v}|S^{\mu}|\mathbf{1}_{v}\rangle=2[\mathbf{2}_{v}|S^{\mu}|\mathbf{1}_{v}].
\end{align}
Thus the two definitions are consistent, with one being the one-particle matrix element of the other.


\section{Three-point amplitude}\label{sec:3Pt}

We study in this section the on-shell three-point amplitudes of HPET.
The main goal here will be to express the most general three-point on-shell amplitude for 
two massive particles (mass $m$, spin $s$) and one massless boson (helicity $h$) in terms of on-shell HPET variables.
Focusing on the minimal coupling portion of such an expression, we will be left with
a resummed form of the HPET three-point amplitude, valid for any spin.
Moreover, we will find that a certain choice of the residual
momentum results in the exponentiation of the minimally coupled three-point amplitude.


In the traditional on-shell variables, the most general three-point amplitude for
two massive particles of mass $m$ and spin $s$, and one massless particle with momentum $q$ and helicity $h$ is \cite{Arkani-Hamed:2017jhn}
\begin{align}
	\mathcal{M}^{+|h|,s} &=(-1)^{2s+h} \frac{x^{|h|}}{m^{2s}}\left[g_0 \langle \mathbf{2}\mathbf{1}\rangle^{2s} 
		+g_1 \langle \mathbf{2}\mathbf{1}\rangle^{2s-1}\frac{x\langle \mathbf{2}q\rangle\langle q
\mathbf{1}\rangle}{m} + \dots + g_{2s}\frac{\left( x\langle\mathbf{2}q\rangle\langle q\mathbf{1}\rangle\right)^{2s}}{m^{2s}}\right],\label{eq:Gen3PtAng} \\
\mathcal{M}^{-|h|,s} &= (-1)^h\frac{x^{-|h|}}{m^{2s}}\left[\tilde{g}_0 [\mathbf{2}\mathbf{1}]^{2s} 
	+\tilde{g}_1 [ \mathbf{2}\mathbf{1}]^{2s-1}\frac{x[ \mathbf{2}q][ q
\mathbf{1}]}{m} + \dots + \tilde{g}_{2s}\frac{\left( x[\mathbf{2}q][ q\mathbf{1}]\right)^{2s}}{m^{2s}}\right].\label{eq:Gen3PtSq}
\end{align}
%
The overall sign differs from the expression in ref.~\cite{Arkani-Hamed:2017jhn}, due to our convention that
$p_1$ is incoming. 
The positive helicity amplitude is expressed in the chiral basis, and the negative helicity amplitude in the
anti-chiral basis.
The minimal coupling portion of this is the amplitude with all couplings except $g_0$ and $\tilde{g}_{0}$ set to zero:
\begin{align}
	\label{eq:3ptMinimalChiral}
	\mathcal{M}^{+|h|,s}_{\textrm{min}} &= (-1)^{2s+h}\frac{g_0 x^{+|h|}}{m^{2s}} \langle \mathbf{2}\mathbf{1}\rangle^{2s}, \\
	\label{eq:3ptMinimalAntichiral}
	\mathcal{M}^{-|h|,s}_{\textrm{min}} &= (-1)^h\frac{\tilde{g}_0 x^{-|h|}}{m^{2s}} [\mathbf{2}\mathbf{1}]^{2s}.
\end{align}
Thus we see that expressing this in terms of on-shell HPET variables
requires that we convert the spinor products $\langle \mathbf{2}\mathbf{1}\rangle$, $x \langle \mathbf{2} q\rangle
\langle q \mathbf{1}\rangle$ (and their anti-chiral basis counterparts) to 
the on-shell HQET variables.

In the remainder of this section we take
$p_{1}^{\mu}=mv^{\mu}+k^{\mu}_{1}$ incoming, and $q^{\mu}$ and $p_{2}^{\mu}=mv^{\mu}+k^{\mu}_{2}$ outgoing. With this choice of kinematics, 
the initial and final residual momenta are related by $k^{\mu}_{2}=k^{\mu}_{1}-q^{\mu}$.
We can relate a spinor with incoming momentum to the spinor with outgoing momentum using analytical continuation,
\cref{eq:AnalCont}. Also, the $x$ factor picks up a negative sign when the directions of $p_{1}$ or $q$ are flipped, $x\rightarrow-x$.


\subsection{General residual momentum}\label{sec:ArbRes}


We start by converting the $s=1/2$ amplitude to on-shell HPET variables.
Inverting \cref{eq:HQETOnShellVars} and simply taking the appropriate spinor products,
we can relate the traditional and HPET spinor products:
%
%
\begin{subequations}\label{eq:OnShellDictAllk}
\begin{align}
    \langle\mathbf{21}\rangle&=\frac{m^{2}}{m_{k_{2}}m_{k_{1}}}\left[\frac{m_{k_{1}}}{m}\langle\mathbf{2}_{v}\mathbf{1}_{v}\rangle+\frac{1}{4m}[\mathbf{2}_{v}q]\langle q\mathbf{1}_{v}\rangle+\frac{x^{-1}}{4m}[\mathbf{2}_{v}q][q\mathbf{1}_{v}]\right], \\
    \langle\mathbf{2}q\rangle\langle q\mathbf{1}\rangle&=\frac{m^{2}}{4m_{k_{2}}m_{k_{1}}}\left(\langle\mathbf{2}_{v}q\rangle\langle q\mathbf{1}_{v}\rangle+x^{-1}\langle\mathbf{2}_{v}q\rangle[ q\mathbf{1}_{v}]+x^{-1}[\mathbf{2}_{v}q]\langle q\mathbf{1}_{v}\rangle+x^{-2}[\mathbf{2}_{v}q][q\mathbf{1}_{v}]\right).
\end{align}
\end{subequations}
%
Similarly, the spinor products in the anti-chiral basis become
\begin{subequations}\label{eq:OnShellDictAllkSq}
\begin{align}
    [\mathbf{2}\mathbf{1}]&=\frac{m^{2}}{m_{k_{2}}m_{k_{1}}}\left[\frac{m_{k_{1}}}{m}[\mathbf{2}_{v}\mathbf{1}_{v}]+\frac{1}{4m}\langle\mathbf{2}_{v}q\rangle[q\mathbf{1}_{v}]+\frac{x}{4m}\langle\mathbf{2}_{v}q\rangle\langle q\mathbf{1}_{v}\rangle\right], \\
    [\mathbf{2}q][q\mathbf{1}]&=\frac{m^{2}}{4m_{k_{2}}m_{k_{1}}}\left([\mathbf{2}_{v}q][q\mathbf{1}_{v}]+x[\mathbf{2}_{v}q]\langle q\mathbf{1}_{v}\rangle+x\langle\mathbf{2}_{v}q\rangle[q\mathbf{1}_{v}]+x^{2}\langle\mathbf{2}_{v}q\rangle\langle q\mathbf{1}_{v}\rangle\right).
\end{align}
\end{subequations}
By substituting \cref{eq:OnShellDictAllkSq,eq:OnShellDictAllk} in \cref{eq:3ptMinimalChiral,eq:3ptMinimalAntichiral}
for $s=1/2$, the minimally coupled amplitudes for positive and negative helicity become
\begin{subequations}
	\label{eq:3ptHPETgenK}
\begin{align}
	\mathcal{M}^{+|h|,s=\frac{1}{2}}_{\text{HPET,min}}&=(-1)^{1+h}g_{0}x^{|h|}\frac{m}{m_{k_{2}}m_{k_{1}}}\left[\frac{m_{k_{1}}}{m}\langle\mathbf{2}_{v}\mathbf{1}_{v}\rangle+\frac{1}{4m}[\mathbf{2}_{v}q]\langle q\mathbf{1}_{v}\rangle+\frac{x^{-1}}{4m}[\mathbf{2}_{v}q][q\mathbf{1}_{v}]\right], \\
	\mathcal{M}^{-|h|,s=\frac{1}{2}}_{\text{HPET,min}}&=(-1)^h\tilde{g}_{0}x^{-|h|}\frac{m}{m_{k_{2}}m_{k_{1}}}\left[\frac{m_{k_{1}}}{m}[\mathbf{2}_{v}\mathbf{1}_{v}]+\frac{1}{4m}\langle\mathbf{2}_{v}q\rangle[q\mathbf{1}_{v}]+\frac{x}{4m}\langle\mathbf{2}_{v}q\rangle\langle q\mathbf{1}_{v}\rangle\right],
\end{align}
\end{subequations}
%
One can expand the $m_{k_{i}}$ in powers of $|k|/m$, which is the characteristic expansion of HPETs.
These three-point amplitudes therefore provide a conjecture for the resummed spin-1/2 HPET amplitude.
Comparing the expansions of \cref{eq:3ptHPETgenK} with that computed directly from the spin-1/2 HPET Lagrangians, 
we have confirmed that they agree at least up to $\mathcal{O}(m^{-2})$ for HQET, and $\mathcal{O}(m^{-1})$ for HBET.\footnote{Note that the power counting of the HBET operators starts one power of $m$ higher than HQET, at $\mathcal{O}(m)$. Thus both of these checks account for the operators up to and including NNLO.} 
Some subtleties of the matching to the Lagrangian calculation are discussed in \Cref{sec:HPETMatch}.

The spin-dependence of these amplitudes can be made 
explicit by using the on-shell form of $q\cdot S$ in \cref{eq:SpinGenRes}:
\begin{subequations}\label{eq:3PtGenkExpSpin}
\begin{align}
	\mathcal{M}^{+|h|,s=\frac{1}{2}}_{\text{HPET,min}}&=(-1)^{1+h}g_{0}x^{|h|}\frac{m}{m_{k_{2}}m_{k_{1}}}\langle\mathbf{2}_{v}|\left[1-\frac{\slashed{v}\slashed{k}_{1}\slashed{k}_{2}\slashed{v}}{4m^{2}}+\frac{q\cdot S}{m}\right]|\mathbf{1}_{v}\rangle, \\
	\mathcal{M}^{-|h|,s=\frac{1}{2}}_{\text{HPET,min}}&=(-1)^h\tilde{g}_{0}x^{-|h|}\frac{m}{m_{k_{2}}m_{k_{1}}}[\mathbf{2}_{v}|\left[1-\frac{\slashed{v}\slashed{k}_{1}\slashed{k}_{2}\slashed{v}}{4m^{2}}-\frac{q\cdot S}{m}\right]|\mathbf{1}_{v}].
\end{align}
\end{subequations}
%
Written in this way, it is immediately apparent how the $k_{1}=0$ parameterization can be obtained from the general case. We turn now to this scenario.

\subsection{Zero initial residual momentum}\label{sec:ZeroRes}

We now consider the parameterization where $k_1^\mu =0$ and $k_2^\mu=-q^\mu$
. 
With zero initial residual momentum, we can switch between the chiral and anti-chiral bases using \cref{eq:HQETMix}:
\begin{subequations}\label{eq:BasisTrans}
\begin{align}
    \langle\mathbf{2}_{v}\mathbf{1}_{v}\rangle&=-[\mathbf{2}_{v}\mathbf{1}_{v}],\label{eq:SpinlessBasisTrans} \\
    \langle\mathbf{2}_{v}q\rangle\langle q\mathbf{1}_{v}\rangle&=x^{-2}[\mathbf{2}_{v}q][q\mathbf{1}_{v}].\label{eq:SpinBasisTrans}
\end{align}
\end{subequations}
Recognizing \cref{eq:SpinlessBasisTrans,eq:SpinBasisTrans} as directly relating spinless effects and the spin-vector respectively in different bases, we see that, for this parameterization, spin effects are never obscured by working in any particular basis.
This is in contrast to the traditional on-shell variables, where the
analog to \cref{eq:SpinlessBasisTrans} includes a spin term, thus hiding or exposing spin dependence
when working in a certain basis.
Thus we have gained a basis-independent interpretation of spinless and spin-inclusive terms.

Either setting $k_{1}=0$ in \cref{eq:3PtGenkExpSpin}, or applying \cref{eq:SpinlessBasisTrans,eq:SpinBasisTrans}
to \cref{eq:OnShellDictAllk,eq:OnShellDictAllkSq}, 
the minimally coupled three-point amplitude with zero residual momentum is obtained:
\begin{subequations}
	\label{eq:3ptMinGensZeroK}
\begin{align}
	\mathcal{M}^{+|h|,s=\frac{1}{2}}_{\text{HPET,min}}&=(-1)^{1+h}\frac{g_{0}x^{|h|}}{m}\left[\langle\mathbf{2}_{v}\mathbf{1}_{v}\rangle+\frac{x}{2m}\langle\mathbf{2}_{v}q\rangle\langle q\mathbf{1}_{v}\rangle\right], \\
	\mathcal{M}^{-|h|,s=\frac{1}{2}}_{\text{HPET,min}}&=(-1)^h\frac{\tilde{g}_{0}x^{-|h|}}{m}\left[[\mathbf{2}_{v}\mathbf{1}_{v}]+\frac{x^{-1}}{2m}[\mathbf{2}_{v}q][q\mathbf{1}_{v}]\right],
\end{align}
\end{subequations}
Note the negative signs which come from treating $p_1$ as incoming.

Three-point kinematics are restrictive enough when $k_{1}=0$ that
we can derive the three-point amplitude in \cref{eq:3ptMinGensZeroK} in an entirely different fashion.
The full three-point amplitude for a heavy spin-$1/2$ particle coupled to a photon can be written as\footnote{We use $\cA$ to denote
a Yang-Mills amplitude.}
\begin{align}
	\mathcal{A}(-\mathbf{1}^{\frac{1}{2}},\mathbf{2}^{\frac{1}{2}},q^{h})=f(m,v,q)ev_{\mu}\epsilon_{q}^{h,\mu}\bar{u}_{v}(p_2)u_{v}(p_1)+g(m,v,q)eq^{\mu}\epsilon_{q}^{h,\nu}\bar{u}_{v}(p_2)\sigma_{\mu\nu}u_{v}(p_1).\label{eq:HQET3pt}
\end{align}
%
The negative in the argument of the amplitude signifies an incoming momentum. 
The three-point operators in the HQET Lagrangian, as well as any non-minimal couplings, modify the functions $f$ and $g$, but there are no other spinor structures that can arise. 
We therefore have two spinor contractions in terms of which we would like to express the spinor brackets of interest. 
We proceed by writing the two contractions in terms of the traditional on-shell variables, and equating this to the contractions expressed in terms of the on-shell HPET variables. 
Working with, say, a positive helicity photon, this yields
\begin{subequations}
\begin{align}
	v_{\mu}\epsilon^{+,\mu}_{q}\bar{u}_{v}(p_2)u_{v}(p_1)&
	=-\sqrt{2}x\langle\mathbf{2}_{v}\mathbf{1}_{v}\rangle
	=-\frac{x}{\sqrt{2}}\left(-\frac{x}{m}\langle\mathbf{2}q\rangle\langle q\mathbf{1}\rangle+2\langle\mathbf{21}\rangle\right)
	,\label{eq:HQETSpinless} \\
	\bar{u}_{v}(p_2)\sigma_{\mu\nu}u_{v}(p_1)q^{\mu}\epsilon^{+,\nu}_{q}&
	=\sqrt{2}ix^{2}\langle\mathbf{2}_{v}q\rangle\langle q\mathbf{1}_{v}\rangle
	=\sqrt{2}ix^{2}\langle\mathbf{2}q\rangle\langle q\mathbf{1}\rangle
	\label{eq:HQETSpin}.
\end{align}
\end{subequations}
Solving for the traditional spinor products, we find
\begin{subequations}\label{eq:OnShellDict}
\begin{align}
    \langle\mathbf{2}\mathbf{1}\rangle&=\langle\mathbf{2}_{v}\mathbf{1}_{v}\rangle+\frac{x}{2m}\langle\mathbf{2}_{v}q\rangle\langle q\mathbf{1}_{v}\rangle,\label{eq:MinCoup} \\
    \langle\mathbf{2}q\rangle\langle q\mathbf{1}\rangle&=\langle\mathbf{2}_{v}q\rangle\langle q\mathbf{1}_{v}\rangle.\label{eq:NonMinCoup}
\end{align}
\end{subequations}
Similarly,
\begin{subequations}\label{eq:OnShellDictSq}
\begin{align}
    [\mathbf{2}\mathbf{1}]&=[\mathbf{2}_{v}\mathbf{1}_{v}]+\frac{x^{-1}}{2m}[\mathbf{2}_{v}q][ q\mathbf{1}_{v}],\label{eq:MinCoupSq} \\
    [\mathbf{2}q][ q\mathbf{1}]&=[\mathbf{2}_{v}q][q\mathbf{1}_{v}].\label{eq:NonMinCoupSq}
\end{align}
\end{subequations}
%
%
Note that \cref{eq:OnShellDict,eq:OnShellDictSq} decompose
the spinor brackets into spinless and spin-inclusive terms. 
Applying \cref{eq:BasisTrans}, it is easy to check that 
this separation of different spin multipoles is independent of the basis used to express
the traditional spinor brackets.

With \cref{eq:HQETSpinless,eq:HQETSpin}, we can rewrite \cref{eq:HQET3pt} as
\begin{align}
    \mathcal{A}(-\mathbf{1}^{\frac{1}{2}},\mathbf{2}^{\frac{1}{2}},q^{+})=\sqrt{2}xe\left(-f(m,v,q)\langle\mathbf{2}_{v}\mathbf{1}_{v}\rangle+g(m,v,q)ix\langle\mathbf{2}_{v}q\rangle\langle q\mathbf{1}_{v}\rangle\right)\label{eq:HQET3pt2}.
\end{align}
The three-point amplitude in QED --- with interaction term $\mathcal{L}_{\text{int}}=e\bar{\psi}\slashed{A}\psi$ --- for a positive helicity photon is
\begin{align}
	\mathcal{A}_{\text{QED}}(-\mathbf{1}^{\frac{1}{2}},\mathbf{2}^{\frac{1}{2}},q^{+})&=e\bar{u}(p_2)\gamma_{\mu}u(p_1)\epsilon^{+,\mu}_{q} \notag \\
    &=\sqrt{2}ex\langle\mathbf{2}\mathbf{1}\rangle,
\end{align}
where in the first line we use Dirac spinors instead of HQET spinors. Substituting \cref{eq:MinCoup} into the above equation gives
\begin{align}
    \mathcal{A}_{\text{QED}}(-\mathbf{1}^{\frac{1}{2}},\mathbf{2}^{\frac{1}{2}},q^{+})&=\sqrt{2}ex\left(\langle\mathbf{2}_{v}\mathbf{1}_{v}\rangle+\frac{x}{2m}\langle\mathbf{2}_{v}q\rangle\langle q\mathbf{1}_{v}\rangle\right).\label{eq:QED3pt}
\end{align}
As abelian HQET is an effective theory derived from QED, it must reproduce the on-shell QED amplitudes when all operators are accounted for. This means that \cref{eq:HQET3pt2,eq:QED3pt} are equal, so we can solve for the functions $f$ and $g$:
\begin{subequations}
\begin{align}
    f(m,v,q)&=-1,\label{eq:HQETf} \\
    g(m,v,q)&=\frac{i}{2m}\label{eq:HQETg}.
\end{align}
\end{subequations}
As a consequence of \cref{eq:HQETf,eq:HQETg}, we conclude that only the leading spin and leading spinless three-point operators of HQET are non-vanishing on-shell when $k_{1}=0$. Indeed, in this case the transfer momentum $q^{\mu}$ is the only parameter that can appear in the HQET operator expansion. In the three-point amplitude, it can only appear in the scalar combinations $q^{2}=0$ by on-shellness of the photon, $v\cdot q\sim q^{2}=0$ by on-shellness of the quarks, or $q\cdot\epsilon(q)=0$ by transversality of the polarization.

To sum up, we list the three-point amplitude for two equal mass spin-1/2 particles and an outgoing photon for both helicities, and in both the chiral and anti-chiral bases:\footnote{We abbreviate the arguments of the amplitude here, but still use $p_{1}$ incoming.}
\begin{subequations}\label{eq:AllHQET3Pt}
\begin{align}
    \cA^{+1,s=\frac{1}{2}}&=\sqrt{2}ex\left(\langle\mathbf{2}_{v}\mathbf{1}_{v}\rangle+\frac{x}{2m}\langle\mathbf{2}_{v}q\rangle\langle q\mathbf{1}_{v}\rangle\right)=-\sqrt{2}ex\left([\mathbf{2}_{v}\mathbf{1}_{v}]-\frac{x^{-1}}{2m}[\mathbf{2}_{v}q][q\mathbf{1}_{v}]\right),\label{eq:PosHQET3Pt} \\
    \cA^{-1,s=\frac{1}{2}}&=\sqrt{2}ex^{-1}\left(\langle\mathbf{2}_{v}\mathbf{1}_{v}\rangle-\frac{x}{2m}\langle\mathbf{2}_{v}q\rangle\langle q\mathbf{1}_{v}\rangle\right)=-\sqrt{2}ex^{-1}\left([\mathbf{2}_{v}\mathbf{1}_{v}]+\frac{x^{-1}}{2m}[\mathbf{2}_{v}q][q\mathbf{1}_{v}]\right),\label{eq:NegHQET3Pt}
\end{align}
\end{subequations}
so $g_{0}=\tilde{g}_{0}=\sqrt{2}em$. When a graviton is emitted instead of a photon, we simply make the replacement $e\rightarrow-\frac{\kappa m}{2\sqrt{2}}$ and square the overall factors of $x$.

We can obtain the amplitude with general initial residual momentum by reparameterizing the states by means of \cref{eq:ReparamTrans}.


\subsection{Most general three-point amplitude}\label{sec:ArbSpin}

Recall the most general three-point amplitude for two massive particles of spin $s$ and mass $m$ and a massless boson with helicity $h$ in the chiral basis, \cref{eq:Gen3PtAng}:
\begin{align}
	\mathcal{M}^{+|h|,s} &=(-1)^{2s+h} \frac{x^{|h|}}{m^{2s}}\left[g_0 \langle \mathbf{2}\mathbf{1}\rangle^{2s} 
		+g_1 \langle \mathbf{2}\mathbf{1}\rangle^{2s-1}\frac{x\langle \mathbf{2}q\rangle\langle q
\mathbf{1}\rangle}{m} + \dots + g_{2s}\frac{\left( x\langle\mathbf{2}q\rangle\langle q\mathbf{1}\rangle\right)^{2s}}{m^{2s}}\right].
\end{align}
%
When expressing \cref{eq:Gen3PtAng} in terms of the on-shell HPET variables,
setting the initial residual momentum to zero, and applying the binomial expansion, we find that 
\begin{subequations}
\begin{align}
	\mathcal{M}^{+|h|,s}_{3}&=(-1)^{2s+h}\frac{x^{|h|}}{m^{2s}}\sum_{k=0}^{2s}g^{\text{H}}_{s,k} \langle\mathbf{2}_{v}\mathbf{1}_{v}\rangle^{2s-k}\left(\frac{x}{2m}\langle\mathbf{2}_{v}q\rangle\langle q\mathbf{1}_{v}\rangle\right)^{k},\label{eq:3ptPosAng}\quad
	g^{\text{H}}_{s,k}=\sum_{i=0}^{k}g_i {{2s-i}\choose{2s-k}}.
\end{align}
We can express this in the anti-chiral basis using \cref{eq:BasisTrans}:
\begin{align}
	\mathcal{M}^{+|h|,s}_{3}&=\frac{x^{|h|}}{m^{2s}}\sum_{k=0}^{2s}g^{\text{H}}_{s,k}(-1)^{k+h} [\mathbf{2}_{v}\mathbf{1}_{v}]^{2s-k}\left(\frac{x^{-1}}{2m}[\mathbf{2}_{v}q][ q\mathbf{1}_{v}]\right)^{k}.\label{eq:3ptPosSq}
\end{align}
\end{subequations}
The $k^{\text{th}}$ spin-multipole can be isolated by choosing the $k^{\text{th}}$ term in the sum. There are $2s+1$ combinations of the spinor brackets in this sum, consistent with the fact that a spin $s$ particle can only probe up to the $2s^{\text{th}}$ spin order term of the spin-multipole expansion. Note also that the coefficient of the spin monopole term is always equal to its value for minimal coupling, making the monopole term universal in any theory.\footnote{This is consistent with the reparameterization invariance of HQET, which fixes the Wilson coefficients of the spinless operators in the HQET Lagrangian up to order $1/m$ \cite{Luke:1992cs}. As argued above, when the initial residual momentum is set to $0$, these are the only operators contributing to the spin monopole.}

The minimal coupling amplitudes are those in \cref{eq:3ptMinimalChiral,eq:3ptMinimalAntichiral},
which correspond to setting $g_{i>0}=0$.
Translating to the on-shell HPET variables, minimal coupling in \cref{eq:3ptPosAng,eq:3ptPosSq} corresponds to $g_{s,k}^{\text{H}}=g_{0}{{2s}\choose{k}}$.

We can write the analogous expressions to \cref{eq:3ptPosAng,eq:3ptPosSq} for a negative helicity massless particle.
Expressing \cref{eq:Gen3PtSq} using \cref{eq:MinCoupSq,eq:NonMinCoupSq},
\begin{subequations}
\begin{align}
	\mathcal{M}^{-|h|,s}_{3}&=(-1)^{h}\frac{x^{-|h|}}{m^{2s-1}}\sum_{k=0}^{2s}\tilde{g}^{\text{H}}_{s,k} [\mathbf{2}_{v}\mathbf{1}_{v}]^{2s-k}\left(\frac{x^{-1}}{2m}[\mathbf{2}_{v}q][q\mathbf{1}_{v}]\right)^{k},\label{eq:3ptNegSq}\quad
	\tilde{g}^{\text{H}}_{s,k}=\sum_{i=0}^{k}\tilde{g}_i {{2s-i}\choose{2s-k}}.
\end{align}
Converting to the chiral basis,
\begin{align}\label{eq:3ptNegAng}
	\mathcal{M}^{-|h|,s}_{3}&=\frac{x^{-|h|}}{m^{2s}}\sum_{k=0}^{2s}\tilde{g}^{\text{H}}_{s,k}(-1)^{2s+h+k} \langle\mathbf{2}_{v}\mathbf{1}_{v}\rangle^{2s-k}\left(\frac{x}{2m}\langle\mathbf{2}_{v}q\rangle\langle q\mathbf{1}_{v}\rangle\right)^{k}.
\end{align}
\end{subequations}
Minimal coupling in this case corresponds to $\tilde{g}_{i>0}=0$, and thus $\tilde{g}^{\text{H}}_{s,k}=\tilde{g}_{0}{{2s}\choose{k}}$.


\subsection{Infinite spin limit}\label{sec:InfSpin}

Various methods have been used to show that the minimal coupling 
three-point amplitude in traditional on-shell variables exponentiates in the infinite spin limit \cite{Guevara:2018wpp,Arkani-Hamed:2019ymq,Guevara:2019fsj}.
All of them require a slight manipulation of the minimal coupling to do so,
with refs.~\cite{Guevara:2018wpp,Guevara:2019fsj} employing 
a change of basis between the chiral and anti-chiral bases, ref.~\cite{Guevara:2018wpp} applying a generalized expectation value,
and refs.~\cite{Guevara:2019fsj,Arkani-Hamed:2019ymq} using a Lorentz boost
-- analogous to the gauge-fixing of the spin operator in ref.~\cite{Levi:2015msa} --
to rewrite the minimal coupling amplitude.
As the on-shell HPET variables inherently make the spin-dependence of the minimal coupling manifest,
the exponentiation of the three-point amplitude is immediate.

Consider the minimal coupling three-point amplitude for two massive spin $s$ particles and one massless particle:
\begin{align}
	\mathcal{M}^{+|h|,s}&=(-1)^{2s+h}\frac{g_{0}x^{|h|}}{m^{2s}}\langle\mathbf{2}_{v}|^{2s}\sum_{k=0}^{2s}\frac{(2s)!}{(2s-k)!}\frac{\left(\frac{x}{2m}|q\rangle\langle q|\right)^{k}}{k!}|\mathbf{1}_{v}\rangle^{2s}.
\end{align}
The quantity in the sum is the rescaled spin-operator $q\cdot S/m$ for a spin $s$ particle, raised to the power of $k$ and divided by $k!$ \cite{Chung:2018kqs},
\begin{align}
    \left(\frac{q\cdot S}{m}\right)^{n}&=\frac{(2s)!}{(2s-n)!}\left(\frac{x}{2m}|q\rangle\langle q|\right)^{n},
\end{align}
where we have suppressed the spinor indices.
The amplitude is therefore
\begin{align}
	\mathcal{M}^{+|h|,s}&=(-1)^{2s+h}\frac{g_{0}x^{|h|}}{m^{2s}}\langle\mathbf{2}_{v}|^{2s}\sum_{k=0}^{2s}\frac{\left(\frac{q\cdot S}{m}\right)^{k}}{k!}|\mathbf{1}_{v}\rangle^{2s}.
\end{align}
We identify the sum with an exponential, with the understanding that the series truncates at the $2s^{\text{th}}$ term for a spin $2s$ particle:
\begin{align}\label{eq:3PtExpAmp}
	\mathcal{M}^{+|h|,s}&=(-1)^{2s+h}\frac{g_{0}x^{|h|}}{m^{2s}}\langle\mathbf{2}_{v}|^{2s}e^{q\cdot S/m}|\mathbf{1}_{v}\rangle^{2s}.
\end{align}
Taking the infinite spin limit, the exponential is exact as its Taylor series does not truncate. We treat the exponential as a number in this limit and remove it from between the spinors \cite{Guevara:2019fsj}:
\begin{align}
	\lim_{s\rightarrow\infty}\mathcal{M}^{+|h|,s}&=\lim_{s\rightarrow\infty}(-1)^{2s+h}\frac{g_{0}x^{|h|}}{m^{2s}}e^{q\cdot S/m}\langle\mathbf{2}_{v}\mathbf{1}_{v}\rangle^{2s}.
\end{align}
Note that since the initial residual momentum is 0, both spinors are associated with the same momentum. 
Then, using the on-shell conditions for these variables,\footnote{The validity of using the on-shell conditions can be checked explicitly 
by rewriting the bracket in terms of traditional on-shell variables, then boosting one of the momenta into the other as in ref.~\cite{Arkani-Hamed:2019ymq}.}
\begin{align}
	\lim_{s\rightarrow\infty}\mathcal{M}^{+|h|,s}&=(-1)^{h}g_{0}x^{|h|}e^{q\cdot S/m}.
\end{align}
This amplitude immediately agrees with the three-point amplitude in refs.~\cite{Guevara:2018wpp,Guevara:2019fsj}: it is the scalar three-point amplitude multiplied by an exponential containing the classical spin-multipole moments. Also notable is that the generalized expectation value (GEV) of ref.~\cite{Guevara:2018wpp} or the Lorentz boosts of refs.~\cite{Arkani-Hamed:2019ymq,Guevara:2019fsj} are not necessary here to interpret the spin dependence classically. 

For the emission of a negative helicity boson, the $n^{\text{th}}$ power of the spin-operator projected along the direction of the boson's momentum is
\begin{align}
    \left(\frac{q\cdot S}{m}\right)^{n}&=\frac{(2s)!}{(2s-n)!}\left(-\frac{x^{-1}}{2m}|q][q|\right)^{n}.
\end{align}
Starting with \cref{eq:3ptNegSq}, the three-point amplitude exponentiates as
\begin{align}\label{eq:ExpNegHel}
	\mathcal{M}^{-|h|,s}&=(-1)^{h}\frac{\tilde{g}_{0}x^{-|h|}}{m^{2s}}[\mathbf{2}_{v}|^{2s}e^{-q\cdot S/m}|\mathbf{1}_{v}]^{2s},
\end{align}
with the exponential being truncated at the $2s^{\textrm{th}}$ term. Taking the infinite spin limit, we find
\begin{align}
	\lim_{s\rightarrow\infty}\mathcal{M}^{-|h|,s}&=\lim_{s\rightarrow\infty}(-1)^{h}\frac{\tilde{g}_{0}x^{-|h|}}{m^{2s}}e^{-q\cdot S/m}[\mathbf{2}_{v}\mathbf{1}_{v}]^{2s}.
\end{align}
Applying the on-shell conditions for these variables, we get 
\begin{align}
	\lim_{s\rightarrow\infty}\mathcal{M}^{-|h|,s}&=(-1)^{h}\tilde{g}_{0}x^{-|h|}e^{-q\cdot S/m}.
\end{align}
Once again we find the scalar three-point amplitude multiplied by an exponential containing the classical spin dependence.

That the exponentials in this section are functions of $q\cdot S$ instead of $2q\cdot S$, as is the case when the traditional on-shell variables are na\"{i}vely exponentiated --- that is, without normalizing by the GEV, or Lorentz boosting one of the spinors --- is significant. We discuss the implications of this in the next section.


\section{Kerr black holes as heavy particles}\label{sec:OnePartEffAct}

In this section, we apply the on-shell HPET variables to the classical
gravitational scattering of two spinning black holes. 
We show that, with the correct momentum parameterization, a heavy spin-$s$ particle minimally
coupled to gravity possesses precisely the same spin-multipole expansion as a Kerr black hole,
up to the order $2s$ multipole.
The reason for this is that on-shell HPET variables for a given velocity $v^{\mu}$,
residual momentum $k^{\mu}$, and mass $m$
always correspond to momenta $m_{k}v^{\mu}$, where $m_{k}$ is defined in \cref{eq:HQETVarMass}.


We begin with a brief review of the effective field theory for spinning gravitating bodies.
The action of a particle interacting with gravitational radiation of wavelength much larger than its spatial extent (approximately a point particle) was formulated in ref.~\cite{Goldberger:2004jt}. 
The generalization to the case of spinning particles was first approached in ref.~\cite{Porto:2005ac}. The effective action formulated in ref.~\cite{Levi:2015msa} takes the form 
\begin{align}\label{eq:OnePartEffAct}
    S&=\int d\sigma \left\{-m\sqrt{u^{2}}-\frac{1}{2}S_{\mu\nu}\Omega^{\mu\nu}+L_{\text{SI}}[u^{\mu},S_{\mu\nu},g_{\mu\nu}(x^{\mu})]\right\},
\end{align}
where $\sigma$ parameterizes the wordline of the particle, $u^{\mu}=\frac{dx^{\mu}}{d\sigma}$ is the coordinate velocity, $S_{\mu\nu}$ is the spin operator, $\Omega^{\mu\nu}$ is the angular velocity, and $L_{\text{SI}}$ contains higher spin-multipoles that are dependent on the inner structure of the particle through non-minimal couplings.

The first two terms in \cref{eq:OnePartEffAct} are the spin monopole and dipole terms, and are universal for spinning bodies with any internal configuration. 
We assign to them respectively the coefficients $C_{S^{0}}=C_{S^{1}}=1$.
From an amplitudes perspective, the universality of the spin-monopole coefficient
can be seen from the on-shell HPET variables since the coefficient of the spin-monopole term in 
\cref{eq:3ptPosAng,eq:3ptNegSq} is always equal to its minimal coupling value. 
The universality of the spin-dipole coefficient was argued in refs.~\cite{Chung:2018kqs,Chung:2019duq} from general covariance, and by requiring the correct factorization of the Compton scattering amplitude. 
Explicitly, the higher spin-multipole terms $L_{\text{SI}}$ are 
\begin{align}
    L_{\text{SI}}&=\sum^{\infty}_{n=1}\frac{(-1)^{n}}{(2n)!}\frac{C_{S^{2n}}}{m^{2n-1}}D_{\mu_{2n}}\dots D_{\mu_{3}}\frac{E_{\mu_{1}\mu_{2}}}{\sqrt{u^{2}}}S^{\mu_{1}}S^{\mu_{2}}\dots S^{\mu_{2n}}\notag \\
    &\quad+\sum^{\infty}_{n=1}\frac{(-1)^{n}}{(2n+1)!}\frac{C_{S^{2n+1}}}{m^{2n}}D_{\mu_{2n+1}}\dots D_{\mu_{3}}\frac{B_{\mu_{1}\mu_{2}}}{\sqrt{u^{2}}}S^{\mu_{1}}S^{\mu_{2}}\dots S^{\mu_{2n+1}}.
\end{align}
See ref.~\cite{Levi:2015msa} for the derivation and formulation of this action. The Wilson coefficients $C_{S^{k}}$ contain the information about the internal structure of the object, with a Kerr black hole being described by $C_{S^{k}}^{\text{Kerr}}=1$ for all $k$.

The three-point amplitude derived from this action was expressed in traditional spinor-helicity variables in refs.~\cite{Chung:2018kqs,Chung:2019duq},
where it was shown that the spin-multipole expansion is necessarily truncated at
order $2s$ when the polarization tensors of spin $s$ particles are used.
By matching this three-point amplitude with the most general form of a three-point amplitude,
it was found there that in the case of minimal coupling one obtains the Wilson coefficients
of a Kerr black hole in the infinite spin limit.
Following their derivation, but using on-shell HPET variables instead, we find (with all momenta incoming)
%
\begin{align}
    \cM^{+2,s}&=\sum_{a+b\leq s}\frac{\kappa mx^{2}}{2 m^{2s}}C_{S^{a+b}}n_{a,b}^{s}\langle\mathbf{2}_{-v}\mathbf{1}_{v}\rangle^{s-a}\left(-x\frac{\langle\mathbf{2}_{-v}q\rangle\langle q\mathbf{1}_{v}\rangle}{2m}\right)^{a}[\mathbf{2}_{-v}\mathbf{1}_{v}]^{s-b}\left(x^{-1}\frac{[\mathbf{2}_{-v}q][q\mathbf{1}_{v}]}{2m}\right)^{b},\notag \\
    n^{s}_{a,b}&\equiv{{s}\choose{a}}{{s}\choose{b}}.
\end{align}
%
As in refs.~\cite{Chung:2018kqs,Chung:2019duq}, we refer to this representation of the amplitude in a form symmetric in the chiral and anti-chiral bases as the polarization basis. Flipping the directions of $p_{2}$ and $q$ (to allow us to directly compare with \cref{eq:3ptPosAng}), then converting the polarization basis to the chiral basis:
\begin{align}
    \cM^{+2,s}&=\frac{x^{2}}{m^{2s}}(-1)^{2s}\sum_{a+b\leq 2s}\frac{\kappa m}{2 }C_{S^{a+b}}n_{a,b}^{s}\langle\mathbf{2}_{v}\mathbf{1}_{v}\rangle^{2s-a-b}\left(\frac{x}{2m}\langle\mathbf{2}_{v}q\rangle\langle q\mathbf{1}_{v}\rangle\right)^{a+b}.
\end{align}
Comparing with \cref{eq:3ptPosAng}, we obtain a one-to-one relation between the coupling constants of both expansions:
\begin{align}
    g_{s,k}^{\text{H}}&=\frac{\kappa m}{2}C_{S^{k}}\sum_{j=0}^{k}n^{s}_{k-j,j}.
\end{align}
Such a one-to-one relation is consistent with the interpretation of \cref{eq:3ptPosAng} as being a spin-multipole expansion. Focusing on the minimal coupling case, we set $g_{i>0}=0$, which means $g_{s,k}^{\text{H}}=g_{0}{{2s}\choose{k}}$. Normalizing $g_{0}=\kappa m/2$, the coefficients of the one-particle effective action for finite spin take the form
\begin{align}\label{eq:KBHMatched}
    C_{S^{k}}^{\text{min}}&={{2s}\choose{k}}\left[\sum_{j=0}^{k}{{s}\choose{k-j}}{{s}\choose{j}}\right]^{-1}=1.
\end{align}
The final equality is the Chu-Vandermonde identity, valid for all $k$. This suggests that the minimal coupling expressed in the on-shell HPET variables produces precisely the multipole moments of a Kerr black hole, even before taking the infinite spin limit.

Using the same matching technique, refs.~\cite{Chung:2018kqs,Chung:2019duq} showed that, when using traditional on-shell variables,
the minimal coupling three-point amplitude for finite spin $s$
corresponded to Wilson coefficients that deviated from those of a Kerr black hole by terms of order $\mathcal{O}(1/s)$. 
Why is it then that the polarization tensors of finite spin HPET possess the same spin-multipole expansion as a Kerr black hole?
Analyzing the matching performed in refs.~\cite{Chung:2018kqs,Chung:2019duq},
the $s$ dependence there arises from the conversion of the polarization basis to the chiral basis.
The reason for this is that new spin contributions arise from this conversion since the chiral
and anti-chiral bases are mixed by two times the spin-operator:
%
\begin{subequations}\label{eq:ChiAntiChiMix}
\begin{align}
    \langle\mathbf{12}\rangle&=-[\mathbf{12}]+\frac{1}{xm}[\mathbf{1}q][q\mathbf{2}], \\
    [\mathbf{12}]&=-\langle\mathbf{12}\rangle+\frac{x}{m}\langle\mathbf{1}q\rangle\langle q\mathbf{2}\rangle.
\end{align}
\end{subequations}
The second terms on the right hand sides of these equations
encode spin effects,
while the first terms were interpreted to be purely spinless.
However, the left hand sides of these equations contradict the latter interpretation;
the spinor brackets $\langle\mathbf{12}\rangle$ and $[\mathbf{12}]$ themselves contain spin effects.
This is the origin of the observed deviation from $C_{S^{k}}^{\text{Kerr}}$: eq.~\eqref{eq:ChiAntiChiMix}, while exposing some spin-dependence,
does not entirely separate the spinless and spin-inclusive effects
encoded in the traditional minimal coupling amplitude.
The result is the matching of an exact spin-multipole expansion on the one-particle
effective action side, to a rough separation of different spin-multipoles
on the amplitude side.

A similar mismatch to Kerr black holes was seen in ref.~\cite{Guevara:2019fsj},
where the minimal coupling amplitude was shown to produce the
spin dependence\footnote{Ref.~\cite{Guevara:2019fsj} worked exclusively with integer spin.
However the only adaptation that must be made to the results therein
when working with half integer spins is the inclusion of a factor of $(-1)^{2s}=-1$.}
\begin{align}\label{eq:ExpSpinDep}
    \langle\mathbf{21}\rangle&=-[\mathbf{2}|e^{2q\cdot S/m}|\mathbf{1}],
\end{align}
where $S^{\mu}$ is the Pauli-Lubanski pseudovector defined with respect to $p_{1}$.
Expanding the exponential and noting that the series terminates after the
spin-dipole term in this case,
it's easy to see the equivalence between this and eq.~\eqref{eq:ChiAntiChiMix}.
The spin-dependence here differs from that of a Kerr black hole
by a factor of two in the exponential \cite{Vines:2017hyw,Guevara:2018wpp}.
Motivated by arguments in ref.~\cite{Levi:2015msa}, an exact match to the Kerr black hole spin multipole expansion was
obtained in ref.~\cite{Guevara:2019fsj} by noting that additional spin contributions are
hidden in the fact that the polarization vectors $[\mathbf{2}|$ and $|\mathbf{1}]$ represent different momenta.
Writing $[\mathbf{2}|$ as a Lorentz boost of $[\mathbf{1}|$, the true spin-dependence of the minimal coupling bracket was manifested:
\begin{align}
    \langle\mathbf{21}\rangle&\sim-[\mathbf{1}|e^{q\cdot S/m}|\mathbf{1}],
\end{align}
up to an operator acting on the little group index of $[\mathbf{1}|$.
The spin-dependence here matches that of a Kerr black hole,
and also matches what has been made explicit in Section~\ref{sec:InfSpin}.
Using a similar Lorentz boost, the authors of ref.~\cite{Arkani-Hamed:2019ymq}
also showed that the minimal coupling bracket indeed contains the spin-dependence of a Kerr black hole.
We see that 
in the absence of a momentum mismatch between the polarization states
used, the full spin-dependence is manifest,
and the multipole expansion of a finite spin $s$ particle corresponds
exactly to that of a Kerr black hole up to $2s^{\text{th}}$ order.

This mismatch of momenta is avoided entirely when using on-shell HPET variables.
Recall that in general the momentum $p_{v}$ represented by on-shell HPET variables is
\begin{align}\label{eq:HPETVarMom}
    p_{v}^{\mu}=m_{k}v^{\mu}.
\end{align}
Working in the case where the initial residual momentum is zero, as in the rest of this section, 
this reduces to simply $mv_{\alpha\dot{\alpha}}$ for the case of $p_{v,1}$.
For $p_{v,2}$, where $p_{2}=p_{1}-q$ and $q$ is the null transfer momentum,
\begin{align}
    p_{v,2}=\left(1-\frac{q^{2}}{4m^{2}}\right)mv^{\mu}=mv^{\mu}.
\end{align}
Consequently, although the initial and final momenta of the massive particle differ by $q$, the degrees of freedom are arranged in such a way that the external states $|\mathbf{1}_{v}\rangle$ and $|\mathbf{2}_{v}\rangle$ are associated with the same momentum. This explains why we have recovered precisely the Wilson coefficients of a Kerr black hole.
We identify this common momentum with that of the Kerr black hole $p_{\text{Kerr}}^{\mu}=mv^{\mu}$.
From the point of view of spinor products, 
\cref{eq:BasisTrans} shows that on-shell HPET variables
provide an unambiguous and basis-independent
interpretation of spinless and spin-inclusive spinor brackets.
Thus, the entire spin dependence of the minimal coupling amplitude is automatically made explicit,
and is isolated from spinless terms.


In the case of $k_{1}\neq0$, the three-term structure of the minimal coupling amplitude spoils its exponentiation.
The matching to the Kerr black hole spin-multipole moments is therefore obscured,
but is recovered in the reparameterization where $k_{1}$ is set to $0$.
This mismatching of the spin-multipole moments can be attributed to the fact that the
polarization tensors for the initial and final states
no longer correspond to the same momentum, since generally $m_{k_{1}-q}\neq m_{k_{1}}$.

A similar matching analysis has recently been performed in ref.~\cite{Chung:2019yfs} 
for the case of Kerr-Newman black holes.
It was also found there that minimal coupling to electromagnetism reproduces
the classical spin multipoles of a Kerr-Newman black hole in the infinite spin limit,
when the matching is performed using traditional on-shell variables.
Repeating their analysis, but using on-shell HPET variables instead,
we find again that the classical multipoles are reproduced exaclty, even for finite spin.

\section{On-shell amplitudes}\label{sec:OnShellAmps}

In this section, we compute electromagnetic and gravitational amplitudes for the scattering
of minimally coupled spin-$s$ particles in on-shell HPET variables using \cref{eq:OnShellDict,eq:OnShellDictSq,eq:OnShellDictAllk,eq:OnShellDictAllkSq}.
Our goal in this section is two-fold: first, we will show how spin effects remain separated from spinless effects, at the order considered in this work, when using on-shell HPET variables.
Second, we will exploit the explicit $\hbar$ dependence of \cref{eq:OnShellDictAllk,eq:OnShellDictAllkSq}
to isolate the classical portions of the computed amplitudes.
Given that the momenta of the on-shell HPET variables always
reduce to the momentum of a Kerr black hole in the classical limit,
we expect to recover the spin-multipoles of a Kerr black hole in this limit.
We show that, at tree-level, the spin dependence of the leading $\hbar$ portions
factorizes into a product of the classical spin-dependence at three-points.
This is simply a consequence of factorization for boson exchange amplitudes (a result that has already been noted in ref.~\cite{Guevara:2019fsj}).
For same-helicity tree-level radiation processes this results from a spin-multipole universality that we will uncover,
and for the opposite helicity Compton amplitude there will be an additional
factor accounting for its non-uniqueness at higher spins.\footnote{We contrast the factorization for radiation processes here with that in ref.~\cite{Guevara:2018wpp}
by noting that the entire quantum amplitude was factorized there,
whereas we show that the factorization holds also for the leading $\hbar$ contribution.}

\subsection{Counting \texorpdfstring{$\hbar$}{h-bar}}\label{sec:hbarCount}

Given that we will be interested in isolating classical effects, we summarize here the rules for restoring the $\hbar$ dependence in the amplitude \cite{Kosower:2018adc}, and adapt these rules to the on-shell variables.

Powers of $\hbar$ are restored in such a way so as to preserve the dimensionality of amplitudes and coupling constants. To do so, the coupling constants of electromagnetism and gravity are rescaled as $e\rightarrow e/\sqrt{\hbar}$ and $\kappa\rightarrow \kappa/\sqrt{\hbar}$. Furthermore, when taking the classical limit $\hbar\rightarrow0$ of momenta, massive momenta and masses are to be kept constant, whereas massless momenta vanish in this limit --- for a massless momentum $q$, it is the associated wave number $\bar{q}=q/\hbar$ that is kept constant in the classical limit. Thus each massless momentum in amplitudes is associated with one power of $\hbar$. Translating this to on-shell variables, we assign a power of $\hbar^{\alpha}$ to each $|q\rangle$, and a power of $\hbar^{1-\alpha}$ to each $|q]$.\footnote{The value of $\alpha$ can be determined by fixing the $\hbar$ scaling of massless polarization tensors for each helicity. Requiring that the dimensions of polarization vectors remain unchanged when $\hbar$ is restored results in the democratic choice $\alpha=1/2$.} Momenta that are treated with the massless $\hbar$ scaling are
\begin{itemize}
    \item photon and graviton momenta, whether they correspond to external or virtual particles;
    \item loop momenta, which can always be assigned to an internal massless boson;
    \item residual momenta \cite{Damgaard:2019lfh}.
\end{itemize}
Finally, we come to the case of spin-inclusive terms. When taking the classical limit $\hbar\rightarrow0$, we simultaneously take the limit $s\rightarrow\infty$ where $s$ is the magnitude of the spin. These limits are to be taken in such a way so as to keep the combination $\hbar s$ constant. This means that for every power of spin in a term, there is one factor of $\hbar$ that we can neglect when taking the classical limit. Effectively, we can simply scale all powers of spin with one inverse power of $\hbar$, and understand that $\hbar$ is to be taken to $0$ wherever it appears in the amplitude.

As in ref.~\cite{Damgaard:2019lfh}, we identify the components of an amplitude 
contributing classically to the interaction potential 
as those with the $\hbar$ scaling
\begin{align}
    \cM\sim\hbar^{-3}.
\end{align}
Terms with more positive powers of $\hbar$ contribute
quantum mechanically to the interaction potential. 
Also, we use $\cM^{\text{cl.}}$ to denote the leading $\hbar$ portion of an amplitude.


\subsection{Boson exchange}\label{sec:tChan}

We begin with the tree-level amplitudes for photon/graviton\footnote{We will denote an amplitude involving photons by $\cA$, and one involving gravitons by $\cM$.} exchange
between two massive spinning particles.
We consider first spin-1/2 -- spin-1/2 scattering,
to show that the spin-multipole expansion remains explicit in these variables at four points.
The classical part of the amplitude can be computed by factorizing it into two three-point amplitudes. 
To simplify the calculation, we are free to set the initial residual momentum of each massive leg to $0$, so we will need only \cref{eq:OnShellDict,eq:OnShellDictSq}. 
Letting particle $a$ have mass $m_{a}$ and incoming/outgoing momenta $p_{1}$/$p_{2}$, and particle $b$ have mass $m_{b}$ and incoming/outgoing momenta $p_{3}$/$p_{4}$, we find for an exchanged photon
\begin{align}
    i\mathcal{A}_{\text{tree}}(-\mathbf{1}^{\frac{1}{2}}_{a},\mathbf{2}^{\frac{1}{2}}_{a},-\mathbf{3}^{\frac{1}{2}}_{b},\mathbf{4}^{\frac{1}{2}}_{b})&=\sum_{h}\mathcal{A}_{\text{tree}}(-\mathbf{1}^{\frac{1}{2}},\mathbf{2}^{\frac{1}{2}},-q^{h})\frac{i}{q^{2}}\mathcal{A}_{\text{tree}}(q^{-h},-\mathbf{3}^{\frac{1}{2}},\mathbf{4}^{\frac{1}{2}})\notag \\
    &=-\frac{ie^{2}}{q^{2}}\left[4\omega\langle\mathbf{2}_{v_{a}}\mathbf{1}_{v_{a}}\rangle\langle\mathbf{4}_{v_{b}}\mathbf{3}_{v_{b}}\rangle\right.\notag \\
    &\quad\left.-\frac{2}{m_{b}}\sqrt{\omega^{2}-1}\ \langle\mathbf{2}_{v_{a}}\mathbf{1}_{v_{a}}\rangle x_{b}\langle\mathbf{4}_{v_{b}}q\rangle\langle q\mathbf{3}_{v_{b}}\rangle\right.\notag \\
    &\quad\left.+\frac{2}{m_{a}}\sqrt{\omega^{2}-1}\ x_{a}\langle\mathbf{2}_{v_{a}}q\rangle\langle q\mathbf{1}_{v_{a}}\rangle\langle\mathbf{4}_{v_{b}}\mathbf{3}_{v_{b}}\rangle\right.\notag \\
    &\quad\left.-\frac{\omega}{m_{a}m_{b}}x_{a}\langle\mathbf{2}_{v_{a}}q\rangle\langle q\mathbf{1}_{v_{a}}\rangle x_{b}\langle\mathbf{4}_{v_{b}}q\rangle\langle q\mathbf{3}_{v_{b}}\rangle\right],\label{eq:EMtChan}
\end{align}
where $\omega\equiv p_{1}\cdot p_{3}/m_{a}m_{b}=(x_{a}x_{b}^{-1}+x_{a}^{-1}x_{b})/2$, $v_{a}=p_{1}/m_{a}$, $v_{b}=p_{3}/m_{b}$, and negative momenta are incoming. The $x$ variables are defined as
\begin{subequations}
\begin{align}
    x_{a}=-\frac{[q|p_{1}|\xi\rangle}{m_{a}\langle q\xi\rangle},&\quad x_{a}^{-1}=-\frac{\langle q|p_{1}|\xi]}{m_{a}[q\xi]}, \\
    x_{b}=\frac{[q|p_{3}|\xi\rangle}{m_{b}\langle q\xi\rangle},&\quad x^{-1}_{b}=\frac{\langle q|p_{3}|\xi]}{m_{b}[q\xi]}.
\end{align}
\end{subequations}
The negative sign in the definitions of $x_{a}$ and $x_{a}^{-1}$ account for the fact that 
the massless boson is incoming to particle $a$.

The gravitational amplitude is computed analogously:
\begin{align}
    i\mathcal{M}_{\text{tree}}(-\mathbf{1}^{\frac{1}{2}}_{a},\mathbf{2}^{\frac{1}{2}}_{a},-\mathbf{3}^{\frac{1}{2}}_{b},\mathbf{4}^{\frac{1}{2}}_{b})&=-\frac{im_{a}m_{b}\kappa^{2}}{8q^{2}}\left[4\left(2\omega^{2}-1\right)\langle\mathbf{2}_{v_{a}}\mathbf{1}_{v_{a}}\rangle \langle\mathbf{4}_{v_{b}}\mathbf{3}_{v_{b}}\rangle\right.\notag \\
    &\quad\left.-\frac{4\omega}{m_{a}}\sqrt{\omega^{2}-1}\ x_{a}\langle\mathbf{2}_{v_{a}}q\rangle\langle q\mathbf{1}_{v_{a}}\rangle\langle\mathbf{4}_{v_{b}}\mathbf{3}_{v_{b}}\rangle\right.\notag \\
    &\quad\left.+\frac{4\omega}{m_{b}}\sqrt{\omega^{2}-1}\ \langle\mathbf{2}_{v_{a}}\mathbf{1}_{v_{a}}\rangle x_{b}\langle\mathbf{4}_{v_{b}}q\rangle\langle q\mathbf{3}_{v_{b}}\rangle\right.\notag \\
    &\quad\left.-\frac{(2\omega^{2}-1)}{m_{a}m_{b}}x_{a}\langle\mathbf{2}_{v_{a}}q\rangle\langle q\mathbf{1}_{v_{a}}\rangle x_{b}\langle\mathbf{4}_{v_{b}}q\rangle\langle q\mathbf{3}_{v_{b}}\rangle\right].\label{eq:GRtChan}
\end{align}
Both amplitudes agree with known results \cite{Holstein:2008sw,Holstein:2008sx,Damgaard:2019lfh}. Furthermore, the amplitudes as written are composed of terms which each individually correspond to a single order in the spin-multipole expansion.
All terms in these amplitudes scale as $\hbar^{-3}$,
so these amplitudes are classical in the sense mentioned in the previous section.

Using the exponential forms of the three-point amplitudes in Section~\ref{sec:InfSpin},
we can write down the boson-exchange amplitudes in the infinite spin case.
We find the same result in the gravitational case as ref.~\cite{Guevara:2019fsj}.
However we have obtained this result immediately simply by gluing together the three-point amplitudes;
we had no need to boost the external states such they represent the same momentum.
Omitting the momentum arguments, the amplitudes are
\begin{subequations}
\begin{align}
    \lim_{s_{a},s_{b}\rightarrow\infty}\cA_{\text{tree}}^{s_{a},s_{b}}&=-\frac{2e^{2}}{q^{2}}\sum_{\pm}(\omega\pm\sqrt{\omega^{2}-1})\ \text{exp}\left[\pm q\cdot\left(\frac{S_{a}}{m_{a}}+\frac{S_{b}}{m_{b}}\right)\right], \\
    \lim_{s_{a},s_{b}\rightarrow\infty}\cM_{\text{tree}}^{s_{a},s_{b}}&=-\frac{\kappa^{2}m_{a}m_{b}}{4q^{2}}\sum_{\pm}(\omega\pm\sqrt{\omega^{2}-1})^{2}\ \text{exp}\left[\pm q\cdot\left(\frac{S_{a}}{m_{a}}+\frac{S_{b}}{m_{b}}\right)\right].
\end{align}
\end{subequations}
The gravitational result corresponds to the first post-Minkowskian (1PM) order amplitude.


\subsection{Compton scattering}\label{sec:Compton}

Our focus shifts now to the electromagnetic and gravitational Compton amplitudes.
These computations will enable the exploitation of the explicit $\hbar$ and spin-multipole expansions to relate the classical limit $\hbar\rightarrow0$ and the classical spin-multipole expansion. 
Concretely, we will show that the spin-multipole expansion of the leading-in-$\hbar$ terms factorizes into a product
of factors of the classical spin-dependence at three-points.

First, consider the spin-$s$ electromagnetic Compton amplitude with two opposite helicity photons,
$\cA(-\mathbf{1}^{s},\mathbf{2}^{s},q_{3}^{-1},q_{4}^{+1})$.
To simplify calculations, we can set the initial residual momentum to $0$, so that $p_{1}^{\mu}=mv^{\mu}$.
Note that it is impossible to set both initial and final residual momenta to $0$ simultaneously, so we will need eqs.~\eqref{eq:OnShellDictAllk} and \eqref{eq:OnShellDictAllkSq}.
We perform the computation by means of Britto-Cachazo-Feng-Witten (BCFW) recursion \cite{Britto:2004ap,Britto:2005fq}, using the $[3,4\rangle$-shift
\begin{subequations}
\begin{align}
    |\hat{4}\rangle&=|4\rangle-z|3\rangle, \quad |\hat{3}]=|3]+z|4].
\end{align}
\end{subequations}
Under this shift, two factorization channels contribute to this amplitude:
\begin{align}
\cA(-\mathbf{1}^{s},\mathbf{2}^{s},q_{3}^{-1},q_{4}^{+1}) &= 
\left.\frac{\cA(-\mathbf{1}^{s},\hat{q}_{3}^{-1},\hat{P}_{13}^{s})\cA(\mathbf{2}^{s},\hat{q}_{4}^{+1},-\hat{P}_{13}^{s})}{\langle 3 | p_1 | 3]}\right|_{\hat{P}_{13}^{2}=m^{2}}\notag \\
&\quad+\left.
\frac{\cA(-\mathbf{1}^{s},\hat{q}_{4}^{+1},\hat{P}_{14}^{s})\cA(\mathbf{2}^{s},\hat{q}_{3}^{-1},-\hat{P}_{14}^{s})}{\langle 4 | p_1 | 4]}\right|_{\hat{P}_{14}^{2}=m^{2}}.
\end{align}
This shift avoids boundary terms for $s\leq1$ as $z\rightarrow\infty$. 
When expressing the factorization channels in terms of on-shell HPET variables,
there is a question about whether new boundary terms arise relative to the traditional
on-shell variables for $z\rightarrow\infty$, as would generally be expected
because of higher-dimensional operators present in EFTs.
This is not the case here, since \cref{eq:spinorInverted} shows that the definition
of the on-shell HPET variables accounts for the contributions from all higher order HPET operators.
Another way to see this is that,
since the relation between the traditional and on-shell HPET variables is exact,
an amplitude must always have the same large $z$ scaling for any shift when
expressed using the on-shell HPET variables as when expressed with the traditional on-shell variables.
Consider for example the spinor contraction part of the $P_{13}$ factorization channel. In the traditional variables, this is
\begin{align}
    \langle\mathbf{2}P_{13}\rangle^{I}{}_{I}[\hat{P}_{13}\mathbf{1}],
\end{align}
which scales as $z$ when $z\rightarrow\infty$. In the on-shell HPET variables:
\begin{align}\label{eq:P13ComptChan}
    \frac{m}{m_{q_{3}+q_{4}}}\left(\langle\mathbf{2}_{v}P_{13v}\rangle_{I}+\frac{1}{4m}[\mathbf{2}_{v}4]\langle\hat{4}P_{13v}\rangle_{I}+\frac{1}{4m\hat{x}_{4}}[\mathbf{2}_{v}4][4\hat{P}_{13v}]_{I}\right)\langle P_{13v}|^{I}\left(\mathbb{I}-\frac{1}{2m\hat{x}_{3}^{-1}}|3\rangle\langle3|\right)|\mathbf{1}_{v}\rangle.
\end{align}
Choosing appropriate reference vectors for $\hat{x}^{-1}_{3}$ and $\hat{x}_{4}$ ($|4]$ and $|3\rangle$ respectively), we recover the unshifted $x^{-1}_{3}$ and $x_{4}$. Thus this also scales as $z$ when $z\rightarrow\infty$. All other factors involved in the factorization channel are common to both sets of variables.

Adding the $P_{13}$ and $P_{14}$ factorization channels, we find the spin-$s$ Compton amplitude
\begin{align}\label{eq:EMCompts}
    \cA(-\mathbf{1}^{s},\mathbf{2}^{s},q_{3}^{-1},q_{4}^{+1})&=(-1)^{2s}\cA(-\mathbf{1}^{0},\mathbf{2}^{0},q_{3}^{-1},q_{4}^{+1})[4|p_{1}|3\rangle^{-2s}\left(1-\frac{q_{3}\cdot q_{4}}{2m^{2}}\right)^{-2s}\notag \\
    &\quad\times\left(\langle3\mathbf{1}_{v}\rangle[4\mathbf{2}_{v}]-\langle3\mathbf{2}_{v}\rangle[4\mathbf{1}_{v}]+\frac{[43]}{2m}\langle\mathbf{2}_{v}3\rangle\langle3\mathbf{1}_{v}\rangle-\frac{\langle34\rangle}{2m}[\mathbf{2}_{v}4][4\mathbf{1}_{v}]\right)^{2s},\notag \\
    \cA(-\mathbf{1}^{0},\mathbf{2}^{0},q_{3}^{-1},q_{4}^{+1})&=-\frac{e^{2}[4|p_{1}|3\rangle^{2}}{\langle4|p_{1}|4]\langle3|p_{1}|3]},
\end{align}
which is in agreement with the result in ref.~\cite{Chung:2018kqs} for QED when the massive spinors are replaced with on-shell HPET spinors.
In the gravitational case, we find
\begin{align}\label{eq:GRCompts}
    \cM(-\mathbf{1}^{s},\mathbf{2}^{s},q_{3}^{-2},q_{4}^{+2})&=(-1)^{2s}\cM(-\mathbf{1}^{0},\mathbf{2}^{0},q_{3}^{-2},q_{4}^{+2})[4|p_{1}|3\rangle^{-2s}\left(1-\frac{q_{3}\cdot q_{4}}{2m^{2}}\right)^{-2s}\notag \\
    &\quad\times\left(\langle3\mathbf{1}_{v}\rangle[4\mathbf{2}_{v}]-\langle3\mathbf{2}_{v}\rangle[4\mathbf{1}_{v}]+\frac{[43]}{2m}\langle\mathbf{2}_{v}3\rangle\langle3\mathbf{1}_{v}\rangle-\frac{\langle34\rangle}{2m}[\mathbf{2}_{v}4][4\mathbf{1}_{v}]\right)^{2s},\notag \\
    \cM(-\mathbf{1}^{0},\mathbf{2}^{0},q_{3}^{-2},q_{4}^{+2})&=-\frac{\kappa^{2}[4|p_{1}|3\rangle^{4}}{8q_{3}\cdot q_{4}\langle4|p_{1}|4]\langle3|p_{1}|3]}.
\end{align}
%
Note the appearance of spurious poles for $s>1$ in the electromagnetic case, 
and for $s>2$ in the gravitational case, 
consistent with the necessarily composite nature of higher spin particles \cite{Arkani-Hamed:2017jhn}. 

Spin effects are isolated in the last two terms in parentheses.
This can be seen in two ways. The first is to rewrite these last two terms in the language of ref.~\cite{Guevara:2018wpp}:
\begin{align}
    \cM(-\mathbf{1}^{s},\mathbf{2}^{s},q_{3}^{-2},q_{4}^{+2})&=\frac{(-1)^{2s}}{m^{2s}}\cM(-\mathbf{1}^{0},\mathbf{2}^{0},q_{3}^{-2},q_{4}^{+2})\left(1-\frac{q_{3}\cdot q_{4}}{2m^{2}}\right)^{-2s}\notag \\
    &\quad\times\langle\mathbf{2}_{v}|^{2s}\left(\mathbb{I}+\frac{1}{2}i\frac{q_{3,\mu}\varepsilon^{-}_{3,\nu}J^{\mu\nu}}{p_{1}\cdot\varepsilon_{3}^{-}}+\frac{1}{2}i\slashed{v}\frac{q_{4,\mu}\varepsilon^{+}_{4,\nu}J^{\mu\nu}}{p_{1}\cdot\varepsilon_{4}^{+}}\slashed{v}\right)^{2s}|\mathbf{1}_{v}\rangle^{2s}.
\end{align}
Alternatively, as is more convenient for our purposes, 
the factorization into classical three-point amplitudes
can be made more visible by application of the Schouten identity to these terms:
\begin{subequations}
\begin{align}\label{eq:GRComptExplicitS}
    \cM(-\mathbf{1}^{s},\mathbf{2}^{s},q_{3}^{-2},q_{4}^{+2})&=\frac{(-1)^{2s}}{m^{2s}}\cM(-\mathbf{1}^{0},\mathbf{2}^{0},q_{3}^{-2},q_{4}^{+2})(\mathcal{N}_{1}+\mathcal{N}_{2})^{2s},
\end{align}
where
\begin{align}
    \mathcal{N}_{1}&\equiv\langle\mathbf{2}_{v}|\left[\mathbb{I}+\frac{(q_{4}-q_{3})\cdot S}{m_{q_{3}+q_{4}}}\right]|\mathbf{1}_{v}\rangle, \\
    \mathcal{N}_{2}&\equiv\langle\mathbf{2}_{v}|\left[v|4]\langle3|\frac{p_{1}\cdot q_{4}}{m_{q_{3}+q_{4}}[4|p_{1}|3\rangle}+|3\rangle[4|v\frac{p_{1}\cdot q_{3}}{m_{q_{3}+q_{4}}[4|p_{1}|3\rangle}\right]|\mathbf{1}_{v}\rangle\notag \\
    &=\langle\mathbf{2}_{v}|\left[v|4]\langle3|\frac{q_{3}\cdot q_{4}}{m_{q_{3}+q_{4}}[4|p_{1}|3\rangle}+\frac{w\cdot S}{m_{q_{3}+q_{4}}}\right]|\mathbf{1}_{v}\rangle,
\end{align}
and
\begin{align}
    w_{\alpha\dot{\alpha}}\equiv2p_{1}\cdot q_{3}\frac{|3\rangle_{\alpha}[4|_{\dot{\alpha}}}{[4|p_{1}|3\rangle},\quad w^{\dot\alpha\alpha}=2p_{1}\cdot q_{3}\frac{|4]^{\dot{\alpha}}\langle3|^{\alpha}}{[4|p_{1}|3\rangle}.
\end{align}
\end{subequations}
$\mathcal{N}_{2}$ is the term that contributes spurious poles for high enough spins. 
The contraction $w\cdot S$ has been defined through \cref{eq:ChiralSpin}.
The momentum $w^{\mu}$ scales linearly with $\hbar$, so the contraction $w\cdot S$ does not scale with $\hbar$.
Compared to this term, the first term in $\mathcal{N}_{2}$ is subleading in $\hbar$. Ignoring it in the classical limit, and
noting that binomial combinatoric factors must be absorbed into the spin-vector when it is raised to some power,
the remaining terms imply an exponential spin structure:
\begin{align}
    \cM^{\text{cl.}}(-\mathbf{1}^{s},\mathbf{2}^{s},q_{3}^{-2},q_{4}^{+2})&=\frac{(-1)^{2s}}{m^{2s}}\cM(-\mathbf{1}^{0},\mathbf{2}^{0},q_{3}^{-2},q_{4}^{+2})\langle\mathbf{2}_{v}|^{2s}\text{exp}\left[\frac{(q_{4}-q_{3}+w)\cdot S}{m}\right]|\mathbf{1}_{v}\rangle^{2s}.
\end{align}
The same exponentiation holds in the electromagnetic case,
with the spinless amplitude above replaced by the corresponding 
spinless amplitude for QED.

The leading $\hbar$ scaling for these amplitudes is $\hbar^{-1}$ whereas na\"{i}ve counting of the vertices and propagators says that the scaling should be $\hbar^{-2}$.
The source of this discrepancy is interference between the two factorization channels, yielding a factor in the numerator of $p_{1}\cdot(\hbar\bar{q}_{3}+\hbar\bar{q}_{4})=\hbar^{2}\bar{q}_{3}\cdot \bar{q}_{4}$.
It is thus possible for the na\"{i}ve $\hbar$ counting to over-count inverse powers of $\hbar$, and hence overestimate the classicality of an amplitude. 
This has consequences for the extension of these results to the emission of $n$ bosons:
factorization channels with a cut graviton line are na\"{i}vely suppressed by one factor of $\hbar$
relative to those with cut matter lines.
The interference described here means that both 
factorizations may actually have the same leading $\hbar$ behavior.

Consider now the same-helicity amplitudes. The two-negative-helicity amplitude for spin-1 has been computed by one of the present authors in ref.~\cite{Aoude:2019tzn} by shifting one massive and one massless leg.
Extending the amplitude found there to spin $s$,
\begin{subequations}
\begin{align}
    \cA(-\mathbf{1}^{s},\mathbf{2}^{s},q_{3}^{-1},q_{4}^{-1})&=\frac{1}{m^{2s}}\cA(-\mathbf{1}^{0},\mathbf{2}^{0},q_{3}^{-1},q_{4}^{-1})[\mathbf{21}]^{2s}, \\
    \cA(-\mathbf{1}^{0},\mathbf{2}^{0},q_{3}^{-1},q_{4}^{-1})&=\frac{e^{2}m^{2}\langle34\rangle^{2}}{\langle3|p_{1}|3]\langle4|p_{1}|4]}
\end{align}
\end{subequations}
We have replaced the coupling in ref.~\cite{Aoude:2019tzn} with $e^{2}$, as is appropriate for QED.
Expressing this in terms of on-shell HPET variables, we find
\begin{align}
    \cA(-\mathbf{1}^{s},\mathbf{2}^{s},q_{3}^{-1},q_{4}^{-1})&=\frac{1}{m^{2s}}\cA(-\mathbf{1}^{0},\mathbf{2}^{0},q_{3}^{-1},q_{4}^{-1})[\mathbf{2}_{v}|^{2s}\left(\mathbb{I}-\frac{(q_{3}+q_{4})\cdot S}{m_{q_{3}+q_{4}}}\right)^{2s}|\mathbf{1}_{v}]^{2s}.
\end{align}
%
The spin-dependence immediately becomes explicit after the change of variables. 
The exponential spin structure is obvious:
\begin{align}
    \cA(-\mathbf{1}^{s},\mathbf{2}^{s},q_{3}^{-1},q_{4}^{-1})&=\frac{1}{m^{2s}}\cA(-\mathbf{1}^{0},\mathbf{2}^{0},q_{3}^{-1},q_{4}^{-1})[\mathbf{2}_{v}|^{2s}\text{exp}\left[-\frac{(q_{3}+q_{4})\cdot S}{m_{q_{3}+q_{4}}}\right]|\mathbf{1}_{v}]^{2s}.
\end{align}

When the gyromagnetic ratio $g=2$, the arbitrary spin $s=s_{1}+s_{2}$ gravitational Compton amplitude
is proportional to the product between the spin $s_{1}$ and $s_{2}$ electromagnetic amplitudes \cite{Choi:1994ax,Bern:2002kj,Bautista:2019tdr}.
As we have constructed the electromagnetic Compton amplitude
using the minimal coupling three-point amplitude, this condition is satisfied.
The same-helicity gravitational Compton amplitude is thus
\begin{subequations}
\begin{align}
    \cM(-\mathbf{1}^{s},\mathbf{2}^{s},q_{3}^{-2},q_{4}^{-2})&=\frac{1}{m^{2s}}\cM(-\mathbf{1}^{0},\mathbf{2}^{0},q_{3}^{-2},q_{4}^{-2})[\mathbf{2}_{v}|^{2s}\text{exp}\left[-\frac{(q_{3}+q_{4})\cdot S}{m_{q_{3}+q_{4}}}\right]|\mathbf{1}_{v}]^{2s}, \\ \cM(-\mathbf{1}^{0},\mathbf{2}^{0},q_{3}^{-2},q_{4}^{-2})&=\frac{\kappa^{2}}{8e^{4}}\frac{\langle3|p_{1}|3]\langle4|p_{1}|4]}{q_{3}\cdot q_{4}}\cA(-\mathbf{1}^{0},\mathbf{2}^{0},q_{3}^{-1},q_{4}^{-1})^{2}.
\end{align}
\end{subequations}
Analogous results hold for the emission of two positive helicity bosons:
\begin{subequations}
\begin{align}
    \cA(-\mathbf{1}^{s},\mathbf{2}^{s},q_{3}^{+1},q_{4}^{+1})&=\frac{1}{m^{2s}}\cA(-\mathbf{1}^{0},\mathbf{2}^{0},q_{3}^{+1},q_{4}^{+1})\langle\mathbf{2}_{v}|^{2s}\text{exp}\left[\frac{(q_{3}+q_{4})\cdot S}{m_{q_{3}+q_{4}}}\right]|\mathbf{1}_{v}\rangle^{2s}, \\
    \cA(-\mathbf{1}^{0},\mathbf{2}^{0},q_{3}^{+1},q_{4}^{+1})&=\frac{e^{2}[34]^{2}}{\langle3|p_{1}|3]\langle4|p_{1}|4]}, \\
    \cM(-\mathbf{1}^{s},\mathbf{2}^{s},q_{3}^{+2},q_{4}^{+2})&=\frac{1}{m^{2s}}\cM(-\mathbf{1}^{0},\mathbf{2}^{0},q_{3}^{+2},q_{4}^{+2})\langle\mathbf{2}_{v}|^{2s}\text{exp}\left[\frac{(q_{3}+q_{4})\cdot S}{m_{q_{3}+q_{4}}}\right]|\mathbf{1}_{v}\rangle^{2s}, \\
    \cM(-\mathbf{1}^{0},\mathbf{2}^{0},q_{3}^{+2},q_{4}^{+2})&=\frac{\kappa^{2}}{8e^{4}}\frac{\langle3|p_{1}|3]\langle4|p_{1}|4]}{q_{3}\cdot q_{4}}\cA(-\mathbf{1}^{0},\mathbf{2}^{0},q_{3}^{+1},q_{4}^{+1})^{2}.
\end{align}
\end{subequations}
Taking the classical limit, we can simply replace $m_{q_{3}+q_{4}}\rightarrow m$
to obtain the leading $\hbar$ behavior of these amplitudes.

To see that the spin-dependence of the leading $\hbar$ portions
of the amplitudes in this section factorize into a product
of the three-point amplitudes, note that
\begin{align}
    {[q_{i}\cdot S,q_{j}\cdot S]_{\alpha}}^{\beta}&=-{\left(v\cdot q_{[i}q_{j]}\cdot S-iq^{\mu}_{i}q^{\nu}_{j}J_{\mu\nu}\right)_{\alpha}}^{\beta}=\mathcal{O}(\hbar),
\end{align}
where square brackets around indices represent normalized anti-symmeterization of the indices.
We can thus combine exponentials and split exponentials of sums
only at the cost of subleading-in-$\hbar$ corrections.

The on-shell HPET variables have made it immediate that the spin exponentiates in the same-helicity Compton amplitudes,
and this exponentiation is preserved in the $\hbar\rightarrow0$ limit.
In the opposite helicity case, the composite nature of higher spin particles can be seen to influence
dynamics already at the leading $\hbar$ level. 
It does so through the contraction $w\cdot S$ for the unphysical momentum $w^{\mu}$,
which appears in a spin exponential in the leading $\hbar$ term.
The focus in this section has been on the emission of two bosons,
but we will now show that the exponentiation in the same-helicity case extends to the $n$ bosons scenario.


\subsection{Emission of \texorpdfstring{$n$}{n} bosons}

We can generalize the exponentiation of the spin observed in the 
same-helicity Compton amplitudes.
In particular, focusing on integer spins for simplicity, we show that for the tree-level
emission of $n$ same-helicity bosons with a common helicity $h$
from a heavy spin-$s$ particle, the amplitude satisfies
\begin{align}\label{eq:nPhotEmis}
    M^{s}_{n+2}&=\frac{(-1)^{nh}}{m^{2s}}M_{n+2}^{s=0}\langle\mathbf{2}_{v}|^{2s}\text{exp}\left[\frac{1}{m_{q}}\frac{h}{|h|}\sum_{i=1}^{n}q_{i}\cdot S\right]|\mathbf{1}_{v}\rangle^{2s} \\
    &=\frac{(-1)^{nh}}{m^{2s}}M_{n+2}^{s=0}[\mathbf{2}_{v}|^{2s}\text{exp}\left[\frac{1}{m_{q}}\frac{h}{|h|}\sum_{i=1}^{n}q_{i}\cdot S\right]|\mathbf{1}_{v}]^{2s}.\notag
\end{align}
We use $q\equiv\sum_{i=1}^{n}q_{i}$ throughout this section.
Once we have proven the first line, the second follows from the fact that the velocity commutes with the spin-vector.
The easiest way to proceed is inductively, constructing the $n+2$ point amplitude
using BCFW recursion.
The cases $n=1,2$ were the focus of previous sections.
Note that the result holds for $n=1$ even when $k_{1}\neq0$, since a non-zero $k_{1}$
results in an additional subleading $\mathcal{O}(\hbar^{2})$ term.

First, note that when expressed in terms of traditional on-shell
variables, the spin dependence in \cref{eq:nPhotEmis} is simply
\begin{subequations}\label{eq:MinCoupSpinDep}
\begin{align}
    \langle\mathbf{21}\rangle^{2s}
    &=\langle\mathbf{2}_{v}|^{2s}\text{exp}\left(\frac{q\cdot S}{m_{q}}\right)|\mathbf{1}_{v}\rangle^{2s}=[\mathbf{2}_{v}|^{2s}\text{exp}\left(\frac{q\cdot S}{m_{q}}\right)|\mathbf{1}_{v}]^{2s},\quad \text{for $h>0$}, \\
    [\mathbf{21}]^{2s}
    &=[\mathbf{2}_{v}|^{2s}\text{exp}\left(-\frac{q\cdot S}{m_{q}}\right)|\mathbf{1}_{v}]^{2s}=\langle\mathbf{2}_{v}|^{2s}\text{exp}\left(-\frac{q\cdot S}{m_{q}}\right)|\mathbf{1}_{v}\rangle^{2s},\quad \text{for $h<0$}.
\end{align}
\end{subequations}
Thus the problem becomes to prove that the spin dependence
is isolated in these spinor contractions.
Having already proven this for the base cases, 
let us now assume it holds up to the emission of $n-1$ bosons
and show that this implies the relations for the emission of $n$ bosons.
Constructing the $n+2$-point amplitude using BCFW, the amplitude takes the general form
\begin{align}\label{eq:nPtBCFW}
    M^{s}_{n+2}&=\sum_{k=1}^{n-1}\sum_{\sigma(k)}\left[\hat{M}^{s,I}_{\sigma(k),k+2}\frac{i\epsilon_{IJ}}{P_{1,\sigma(k)}^{2}}\hat{M}_{\sigma(n-k),n-k+2}^{s,J}+\sum_{h=\pm}\hat{M}^{s,h}_{\sigma(k),k+3}\frac{i}{P^{2}_{0,\sigma(k)}}\hat{M}^{-h}_{\sigma(n-k),n-k+1}\right],
\end{align}
where $P_{1,\sigma(k)}\equiv p_{1}+\sum_{i=1}^{k}q_{\rho(i,\sigma(k))}\equiv p_{1}+P_{0,\sigma(k)}$. The permutations $\sigma(k)$ and $\sigma(n-k)$ account for all the ways
of organizing the boson legs into $k+2$ and $n-k+2$ point amplitudes,
in which shifted legs are never in the same sub-amplitude.
$\rho(i,\sigma(k))$ denotes the $i^{\text{th}}$ index in the permutation $\sigma(k)$.
The notation $\hat{M}$ reminds us that the sub-amplitudes are functions of shifted momenta. 
The first term in \cref{eq:nPtBCFW} represents factorizations where a massive propagator is on-shell,
whereas the second accounts for a massless propagator going on-shell --- $h$ in this second term is the helicity of the cut boson.

We will treat each term in \cref{eq:nPtBCFW} separately. 
We begin with the first term, which is the only contribution for QED.
For the case of $n$ positive-helicity bosons, we shift $|\mathbf{1}]$ and, say, $|q_{1}\rangle$ as in ref.~\cite{Aoude:2019tzn}. Then, applying the induction hypothesis, this term is
\begin{align}
    \frac{(-1)^{nh}}{m^{4s}}&\sum_{k=1}^{n-1}\sum_{\sigma(k)}\hat{M}^{s=0,I}_{\sigma(k),k+2}\frac{i}{P_{1,\sigma(k)}^{2}}\hat{M}_{\sigma(n-k),n-k+2}^{s=0,J}\langle\mathbf{2}\hat{P}_{1,\sigma(k)}^{I}\rangle^{2s}\langle\hat{ P}_{1,\sigma(k)I}\mathbf{1}\rangle^{2s}\notag \\
    &=\frac{(-1)^{nh}}{m^{2s}}\langle\mathbf{2}\mathbf{1}\rangle^{2s}\sum_{k=1}^{n-1}\sum_{\sigma(k)}\hat{M}^{s=0,I}_{\sigma(k),k+2}\frac{i}{P_{1,\sigma(k)}^{2}}\hat{M}_{\sigma(n-k),n-k+2}^{s=0,J}
\end{align}
The case of $n$ negative-helicity bosons can be shown similarly by shifting
$|\mathbf{1}\rangle$ and, say, $|q_{1}]$.
In particular, choosing an appropriate shift of one massive and one massless leg
results in no massive shift appearing in the sub-amplitudes.
Applying \cref{eq:MinCoupSpinDep} to this, the form of the first term in \cref{eq:nPtBCFW} is therefore
\begin{align}\label{eq:nPtBCFWI}
    \frac{(-1)^{nh}}{m^{2s}}\langle\mathbf{2}_{v}|^{2s}\text{exp}\left[\frac{1}{m_{q}}\frac{h}{|h|}\sum_{i=1}^{n}q_{i}\cdot S\right]|\mathbf{1}_{v}\rangle^{2s}\sum_{k=1}^{n-1}\sum_{\sigma(k)}\hat{M}^{s=0}_{\sigma(k),k+2}\frac{i}{P_{1,\sigma(k)}^{2}}\hat{M}^{s=0}_{\sigma(n-k),n-k+2}
\end{align}
The remaining sum here is the BCFW
form of the amplitude for $n$-photon emission from a massive scalar.
Thus we have proven \cref{eq:nPhotEmis} for the photon case.

The non-linear nature of gravity allows contributions from the second term in \cref{eq:nPtBCFW}.
The contribution of this term to the amplitude is predictable for unique-helicity configurations. 
The only non-vanishing factorization channels will involve the product of $(n-1)+2$
point amplitudes with $n-1$ same-helicity gravitons, and 
a three-graviton amplitude with one distinct helicity graviton, which is the cut graviton.
For example, consider the all-plus helicity amplitude.
Applying the induction hypothesis,
\begin{align}\label{eq:nPtBCFWII}
    \sum_{k=1}^{n-1}\sum_{\sigma(k)}&\left.\sum_{h=\pm}\hat{\cM}^{s,h}_{\sigma(k),k+3}\frac{i}{P^{2}_{0,\sigma(k)}}\hat{\cM}^{-h}_{\sigma(n-k),n-k+1}\right|_{\text{cl.}}=\left.\sum_{\sigma(n-2)}\hat{\cM}^{s,+}_{\sigma(n-2),n+1}\frac{i}{P^{2}_{0,\sigma(n-2)}}\hat{\cM}^{-}_{\sigma(2),3}\right|_{\text{cl.}}\notag \\
    &=\frac{1}{m^{2s}}\langle\mathbf{2}_{v}|^{2s}\text{exp}\left[\frac{1}{m_{q}}\sum_{i=1}^{n}q_{i}\cdot S\right]|\mathbf{1}_{v}\rangle^{2s}\sum_{\sigma(n-2)}\hat{\cM}^{s=0,+}_{\sigma(n-2),n+1}\frac{i}{P^{2}_{0,\sigma(n-2)}}\hat{\cM}^{-}_{\sigma(2),3}.
\end{align}
We have used momentum conservation to write the cut momentum in terms of
the sum of the momenta of the gravitons in the all-graviton subamplitude.
The argument is identical in the all-negative case.
Adding \cref{eq:nPtBCFWI,eq:nPtBCFWII} and identifying the remaining sums of sub-amplitudes
as the scalar amplitude for the emission of $n+2$ gravitons, we find
\begin{align}\label{eq:nGravEmis}
    \cM^{s}_{n+2}&=\frac{1}{m^{2s}}\cM_{n+2}^{s=0}\langle\mathbf{2}_{v}|^{2s}\text{exp}\left[\frac{1}{m_{q}}\frac{h}{|h|}\sum_{i=1}^{n}q_{i}\cdot S\right]|\mathbf{1}_{v}\rangle^{2s} \\
    &\quad=\frac{1}{m^{2s}}\cM_{n+2}^{s=0}[\mathbf{2}_{v}|^{2s}\text{exp}\left[\frac{1}{m_{q}}\frac{h}{|h|}\sum_{i=1}^{n}q_{i}\cdot S\right]|\mathbf{1}_{v}]^{2s}.\notag
\end{align}

In amplitudes where this spin universality is manifest,
we can eliminate the dependence on the specific states used by 
taking the infinite spin and classical limits of the result,
\begin{align}
    \lim_{\substack{s\to\infty \\ \hbar\to0}}M^{s}_{n+2}&=M_{n+2}^{s=0}\ \text{exp}\left[\frac{1}{m}\frac{h}{|h|}\sum_{i=1}^{n}q_{i}\cdot S\right],
\end{align}
where we have used that $\lim_{\hbar\rightarrow0}p_{v,2}^{\mu}=\lim_{\hbar\rightarrow0}p_{v,1}^{\mu}=mv^{\mu}$
to apply on-shell conditions.
This makes contact between the classical limit of the kinematics, and the classical spin limit:
for tree-level same-helicity boson emission processes,
the spin dependence of the leading-in-$\hbar$ term 
factorizes into factors of the classical three-point spin-dependence.

\section{Summary and outlook}\label{sec:Conclusion}

We have presented an on-shell formulation of HPETs by expressing their asymptotic states
as a linear combination of the chiral and anti-chiral massive on-shell helicity variables of ref.~\cite{Arkani-Hamed:2017jhn}.
This expression automatically takes into account the infinite tower of higher-dimensional
operators present in HPETs, which result from the integrating out of the anti-field.
The variables defined in this manner possess manifest spin multipole and $\hbar$ expansions.
Consequently, using the most general three-point amplitude of ref.~\cite{Arkani-Hamed:2017jhn},
we have been able to derive a closed form for the amplitude arising
from the sum of all three-point operators in an arbitrary spin HPET.
This form of the amplitude has been checked explicitly up to NNLO in the operator expansion of spin-1/2 HQET and HBET.
We will also show in \Cref{sec:HPETMatch} that the extension to higher spins
is suitable for describing the three-point amplitude for zero initial residual momentum
for a heavy spin-1 particle coupled to electromagnetism.

We have shown that the spin-multipole expansion of minimally coupled 
heavy particles corresponds exactly to a truncated Kerr black hole expansion when the initial residual momentum
is set to zero.
This has been done in two ways. First, we exponentiated the spin dependence of the minimally coupled
three-point amplitude in \Cref{sec:InfSpin}.
Doing so directly produced the same spin exponential as that in 
refs.~\cite{Vines:2017hyw,Guevara:2018wpp}
for a Kerr black hole coupled to a graviton.
Unlike previous approaches, no further manipulation of the three-point amplitude was needed
to match to refs.~\cite{Vines:2017hyw,Guevara:2018wpp}. An exact match to all spin orders was achieved in the infinite spin limit.
An alternative approach to matching the Kerr black hole multipole moments was carried out in refs.~\cite{Chung:2018kqs,Chung:2019duq},
by matching to the EFT of ref.~\cite{Levi:2015msa}.
Following this matching procedure but using on-shell HPET variables,
an exact match to the Kerr black hole Wilson coefficients
was achieved without the need to take an infinite spin limit. 
The reason that the three-point amplitude in on-shell HPET variables
immediately matches the Kerr black hole multipole expansion is that
the heavy spinors representing the initial and final states are both associated with the same momentum,
which is identified with that of the black hole.

We set out to provide a framework that would enable
the extension of HPETs to higher spins, and to enable
the application of HPETs to the computation of higher order
classical amplitudes.
As a step in this direction, we applied recursion relations to
the minimal coupling amplitude for heavy particles to build arbitrary-spin higher-point tree amplitudes.
Doing so, we showed that the explicit $\hbar$ and spin multipole expansions at
three points remained manifest in all amplitudes considered.
We also easily constructed the tree-level boson exchange amplitude to all orders in spin for QED and GR,
without having to further manipulate the states to produce the correct classical black hole spin multipole expansion.

Moving on to radiative processes, we showed that the same-helicity
electromagnetic and gravitational Compton amplitudes exhibit a spin universality: they can be written as
\begin{align}
    M^{s}_{4}&=M^{s=0}_{4}\langle\mathbf{2}_{v}|^{2s} \text{exp}\left[\frac{1}{m_{q_{1}+q_{2}}}\frac{h}{|h|}\sum_{i=1}^{2}q_{i}\cdot S\right]|\mathbf{1}_{v}\rangle^{2s}.
\end{align}
This universality extends to the emission of 
$n$ same-helicity bosons (\cref{eq:nPhotEmis}).
In the four-point opposite-helicity case, a similar exponential was obtained only in the classical limit.
However the sum in the exponential also included an unphysical momentum contracted with the spin,
representing the non-uniqueness of the amplitude for large enough spins.
It would be interesting to examine whether the opposite-helicity amplitude
possesses an $n$-boson extension analogous to \cref{eq:nPhotEmis}.
Another natural extension is to study how the leading $\hbar$ behaviour
changes when a second matter line is included in radiation processes; this is relevant to the understanding of 
non-conservative effects in spinning binaries.
The understanding of radiative processes is paramount to the PM amplitude program,
as the construction of higher PM amplitudes using unitarity methods requires knowledge of tree-level radiative amplitudes.
Combining radiative amplitudes with the $\hbar$ counting of the on-shell HPET variables in a unitarity-based approach,
the classical limits of amplitudes can be easily identified and taken before integration
to simplify computations of classical loop amplitudes including spin.

Because of the topicality of the subject, we have focused in the main body of this paper on
the application of these variables to their interpretation as spinning black holes
and the construction of classical tree-level amplitudes.
Nevertheless, they are equally applicable to the QCD systems which HQET was formulated to describe. 
Moreover an on-shell perspective is useful for the understanding of HPETs as a whole.
Indeed, we take an on-shell approach in the appendices to make further statements about HPETs.

\acknowledgments

We thank Paolo Benincasa, N. E. J. Bjerrum-Bohr, Andrea Cristofoli, Poul H. Damgaard, Jung-Wook Kim, Mich\`{e}le Levi, Alexander Ochirov, and Justin Vines for useful discussions.
We also thank Mich\`{e}le Levi for comments on the manuscript.
This project has received funding from the European Union's Horizon 2020 
research and innovation programme under the Marie Sk\l{}odowska-Curie grant 
agreement No. 764850 "SAGEX".
The work of A.H. was supported in part by the Danish National Research Foundation (DNRF91) and
the Carlsberg Foundation. The work of R.A. was supported
by the Alexander von Humboldt Foundation, in the framework of the Sofja Kovalevskaja Award 2016, endowed by the German Federal Ministry of Education and Research and also supported by  the  Cluster  of  Excellence  ``Precision  Physics,  Fundamental
Interactions, and Structure of Matter" (PRISMA$^+$ EXC 2118/1) funded by the German Research Foundation (DFG) within the German Excellence Strategy (Project ID 39083149).


\appendix


\section{Conventions}\label{sec:Conventions}

We list here our conventions for reference. In the Weyl basis, the Dirac gamma matrices take the explicit form
\begin{align}
    \gamma^{\mu}&=\begin{pmatrix}
    0 & (\sigma^{\mu})_{\alpha\dot{\alpha}} \\
    (\bar{\sigma}^{\mu})^{\dot{\alpha}\alpha} & 0
    \end{pmatrix},
\end{align}
where $\sigma^{\mu}=(1,\sigma^{i})$, $\bar{\sigma}^{\mu}=(1,-\sigma^{i})$, and $\sigma^{i}$ are the Pauli matrices. The gamma matrices obey the Clifford algebra $\{\gamma^{\mu},\gamma^{\nu}\}=2\eta^{\mu\nu}$. We use the mostly minus metric convention, $\eta^{\mu\nu}=\text{diag}\{+,-,-,-\}$. The fifth gamma matrix is defined as
\begin{align}
    \gamma_{5}\equiv i\gamma^{0}\gamma^{1}\gamma^{2}\gamma^{3}=\begin{pmatrix}
    -\mathbb{I} & \ 0 \\
    0 & \ \mathbb{I}
    \end{pmatrix}.
\end{align}
The generator of Lorentz transforms is
\begin{align}
    J^{\mu\nu}=\frac{i}{4}[\gamma^{\mu},\gamma^{\nu}].
\end{align}

We express massless momenta in terms of on-shell variables:
\begin{subequations}
\begin{align}
    q_{\alpha\dot{\alpha}}\equiv q^{\mu}(\sigma_{\mu})_{\alpha\dot{\alpha}}&=\lambda_{\alpha}\tilde{\lambda}_{\dot{\alpha}}\equiv|\lambda\rangle_{\alpha}[\lambda|_{\dot{\alpha}}, \\
    q^{\dot{\alpha}\alpha}\equiv q^{\mu}(\sigma_{\mu})^{\dot{\alpha}\alpha}&=\tilde{\lambda}^{\dot{\alpha}}\lambda^{\alpha}\equiv|\lambda]^{\dot{\alpha}}\langle\lambda|^{\alpha}.
\end{align}
\end{subequations}
Here $\alpha$, $\dot{\alpha}$ are $SL(2,\mathbb{C})$ spinor indices. Spinor brackets are formed by contracting the spinor indices,
\begin{align}
    \langle\lambda_{1}\lambda_{2}\rangle&\equiv\langle\lambda_{1}|^{\alpha}|\lambda_{2}\rangle_{\alpha}, \\
    [\lambda_{1}\lambda_{2}]&\equiv[\lambda_{1}|_{\dot{\alpha}}|\lambda_{2}]^{\dot{\alpha}}.
\end{align}
For massive momenta, we have that 
\begin{subequations}\label{eq:MassiveMom}
\begin{align}
    p_{\alpha\dot{\alpha}}&={\lambda_{\alpha}}^{I}\tilde{\lambda}_{\dot{\alpha}I}\equiv|\lambda\rangle_{\alpha}^{I}[\lambda|_{\dot{\alpha}I}, \\
    p^{\dot{\alpha}\alpha}&={\tilde{\lambda}^{\dot{\alpha}}}_{I}\lambda^{\alpha I}\equiv|\lambda]^{\dot{\alpha}}_{I}\langle\lambda|^{\alpha I},
\end{align}
\end{subequations}
where $I$ is an $SU(2)$ little group index. Spinor brackets for massive momenta are also formed by contracting spinor indices, identically to the massless case. We also use the bold notation introduced in ref.~\cite{Arkani-Hamed:2017jhn} to suppress the symmetrization over $SU(2)$ indices in amplitudes:
\begin{align}
    \langle\mathbf{2}q_{1}\rangle\langle\mathbf{2}q_{2}\rangle&\equiv\begin{cases}
    \langle2^{I}q_{1}\rangle\langle2^{J}q_{2}\rangle & I=J, \\
    \langle2^{I}q_{1}\rangle\langle2^{J}q_{2}\rangle+\langle2^{J}q_{1}\rangle\langle2^{I}q_{2}\rangle & I\neq J.
    \end{cases}
\end{align}

The Levi-Civita symbol, used to raise and lower spinor and $SU(2)$ little group indices, is defined by
\begin{align}
    \epsilon^{12}=-\epsilon_{12}=1.
\end{align}
Spinor and $SU(2)$ indices are raised and lowered by contracting with the second index on the Levi-Civita symbol. For example,
\begin{align}
    \lambda^{I}=\epsilon^{IJ}\lambda_{J},&\quad \lambda_{I}=\epsilon_{IJ}\lambda^{J}.
\end{align}
The on-shell conditions for the massive helicity variables are
\begin{subequations}\label{eq:OnShellness}
\begin{align}
    \lambda^{\alpha I}\lambda_{\alpha J}=m{\delta^{I}}_{J},\quad \lambda^{\alpha I}{\lambda_{\alpha}}^{J}&=-m\epsilon^{IJ},\quad {\lambda^{\alpha}}_{I}\lambda_{\alpha J}=m\epsilon_{IJ}, \\
    {\tilde{\lambda}_{\dot{\alpha}}}^{I}{\tilde{\lambda}^{\dot{\alpha}}}_{J}=-m{\delta^{I}}_{J},\quad {\tilde{\lambda}_{\dot{\alpha}}}^{I}\tilde{\lambda}^{\dot{\alpha} J}&=m\epsilon^{IJ},\quad \tilde{\lambda}_{\dot{\alpha} I}{\tilde{\lambda}^{\dot{\alpha}}}_{J}=-m\epsilon_{IJ}.
\end{align}
\end{subequations}

Given \cref{eq:HQETOnShellVars}, we can derive the on-shell conditions of the HPET variables, analogous to \cref{eq:OnShellness}. We find
\begin{subequations}\label{eq:HQETOnShellness}
\begin{align}
    \lambda^{\alpha I}_{v}\lambda_{v\alpha J}=m_{k}{\delta^{I}}_{J},\quad \lambda^{\alpha I}_{v}{\lambda_{v\alpha}}^{J}&=-m_{k}\epsilon^{IJ},\quad \lambda^{\alpha}_{vI}\lambda_{v\alpha J}=m_{k}\epsilon_{IJ}, \\
    \tilde{\lambda}_{v\dot{\alpha}}^{I}\tilde{\lambda}_{vJ}^{\dot{\alpha}}=-m_{k}{\delta^{I}}_{J},\quad \tilde{\lambda}_{v\dot{\alpha}}^{I}\tilde{\lambda}_{v}^{\dot{\alpha}J}&=m_{k}\epsilon^{IJ},\quad \tilde{\lambda}_{v\dot{\alpha}I}\tilde{\lambda}_{vJ}^{\dot{\alpha}}=-m_{k}\epsilon_{IJ},
\end{align}
where
\begin{align}
    m_{k}&\equiv\left(1-\frac{k^{2}}{4m^{2}}\right)m.
\end{align}
\end{subequations}

In \Cref{sec:RepLG} we will decompose massive momenta into two massless momenta, as in \cref{eq:LCD}. When identifying
\begin{subequations}
\begin{align}
    {\lambda_{\alpha}}^{1}=|a\rangle_{\alpha},&\quad{\lambda_{\alpha}}^{2}=|b\rangle_{\alpha}, \\
    \tilde{\lambda}_{\dot{\alpha}1}=[a|_{\dot{\alpha}},&\quad\tilde{\lambda}_{\dot{\alpha}2}=[b|_{\dot{\alpha}},
\end{align}
\end{subequations}
we use $\langle ba\rangle=[ab]=m$.

On-shell variables can be assigned to the upper and lower Weyl components of a Dirac spinor so that the spinors satisfy the Dirac equation \cite{Chung:2018kqs},
\begin{align}
    u^{I}(p)=\begin{pmatrix}
    {\lambda_{\alpha}}^{I} \\
    \tilde{\lambda}^{\dot{\alpha} I}
    \end{pmatrix},&\quad \bar{u}_{I}(p)=\begin{pmatrix}
    -{\lambda^{\alpha}}_{I} &    \tilde{\lambda}_{\dot{\alpha} I}
    \end{pmatrix},
\end{align}
where $p$ is expressed in terms of $\lambda$ and $\tilde{\lambda}$ as in \cref{eq:MassiveMom}.

Using analytic continuation, under a sign flip of the momentum, the on-shell variables transform as
\begin{subequations}\label{eq:AnalCont}
\begin{align}
    |-\mathbf{p}\rangle=-|\mathbf{p}\rangle,&\quad |-\mathbf{p}]=|\mathbf{p}],
\end{align}
%
which means
\begin{align}
    |-\mathbf{p}_{v}\rangle = |\mathbf{p}_{-v}\rangle=-|\mathbf{p}_{v}\rangle,
    &\quad |-\mathbf{p}_{v}]=|\mathbf{p}_{-v}]=|\mathbf{p}_{v}].
\end{align}
\end{subequations}
%



\section{Uniqueness of on-shell HPET variables}\label{sec:Uniqueness}

In this section, we address the question of uniqueness of the on-shell HPET variables as defined in \cref{eq:HQETOnShellVars}. 
In particular, we 
%
relate the on-shell HPET variables $|\mathbf{p}_{v}\rangle$ and $|\mathbf{p}_{v}]$ 
to the traditional on-shell variables under two conditions:
\begin{enumerate}
    \item\label{Cond:1} The new variables describe a very massive spin-1/2 state that acts as a source for mediating bosons, meaning that the velocity of the state is approximately constant. 
    Since the motion of the particle is always very closely approximated by its velocity, we demand that the new variables satisfy the Dirac equation for a velocity $v^{\mu}$ and mass $v^{2}=1$:
    \begin{align}
        \slashed{v}|\mathbf{p}_{v}\rangle=|\mathbf{p}_{v}],\quad \slashed{v}|\mathbf{p}_{v}]=|\mathbf{p}_{v}\rangle.
    \end{align}
    Clearly these relations can be scaled to give the state an arbitrary mass.
    
    \item\label{Cond:2} When describing a heavy particle with mass $m$ and velocity $v^{\mu}$, 
    the new variables must reduce to the traditional on-shell variables with $p^{\mu}=mv^{\mu}$ when $k=0$.
\end{enumerate}

We express the on-shell HPET variables in the basis of traditional on-shell variables:
\begin{subequations}\label{eq:BotUpVars}
\begin{align}
    |\mathbf{p}_{v}\rangle&=a(k)|\mathbf{p}\rangle+\slashed{\Gamma}_{1}(k)|\mathbf{p}]\label{eq:BotUpVarsAng}, \\
    |\mathbf{p}_{v}]&=b(k)|\mathbf{p}]+\slashed{\Gamma}_{2}(k)|\mathbf{p}\rangle\label{eq:BotUpVarsSq}.
\end{align}
\end{subequations}
The fact that the functions $a,\ b,\ \Gamma_{1},\ \Gamma_{2}$ can, without loss of generality, be assumed to be functions of only $k^{\mu}$ (and $m$) follows from on-shellness and the Dirac equation. Any dependence on $v^{\mu}$ must be either in a scalar form, $v\cdot v=1$ or $v\cdot k=-k^{2}/2m$, or in matrix form $\slashed{v}$, which can be eliminated for $\slashed{k}/m$ using the Dirac equation for $\slashed{p}$. This also means that we can rewrite $\Gamma^{\mu}_{1,2}=c_{1,2}(k)k^{\mu}$, where the $c_{i}(k)$ are scalars and potentially functions of $k^{2}$. 
Moreover, given that $a$ and $b$ are functions only of $k$, they must also be scalars; the only possible matrix combinations they can contain to preserve the correct spinor indices are even powers of $\slashed{k}$, which would reduce to some power of $k^{2}$. Condition~\ref{Cond:2} provides a final constraint on these four functions:
\begin{subequations}\label{eq:CoeffConds}
\begin{align}
    a(0)&=b(0)=1, \\
    \Gamma_{1}(0)&=\Gamma_{2}(0)=0.
\end{align}
\end{subequations}
Since $\Gamma_{i}^{\mu}=c_{i}(k)k^{\mu}$, the second line imposes that the $c_{i}(k)$ are regular at $k=0$. From now on we drop the arguments of these functions for brevity.

Applying condition~\ref{Cond:1} to eqs.~\eqref{eq:BotUpVars}, we derive relations among the four functions $a,\ b,\ c_{1},\ c_{2}$:
\begin{subequations}
\begin{align}
    b&=a, \\
    c_{2}&=-\frac{a}{m}-c_{1}.
\end{align}
\end{subequations}
The most general on-shell HPET variables are thus
\begin{subequations}
\begin{align}
    |\mathbf{p}_{v}\rangle&=a|\mathbf{p}\rangle+c_{1}\slashed{k}|\mathbf{p}]\label{eq:BotUpVarsAngInter}, \\
    |\mathbf{p}_{v}]&=a|\mathbf{p}]-\left(\frac{a}{m}+c_{1}\right)\slashed{k}|\mathbf{p}\rangle\label{eq:BotUpVarsSqInter}.
\end{align}
\end{subequations}
The momentum associated with these states is
\begin{align}
    \slashed{p}_{v}&=\begin{pmatrix}
    0 & |p_{v}\rangle^{I}{}_{I}[p_{v}| \\
    |p_{v}]_{I}{}^{I}\langle p_{v}| & 0
    \end{pmatrix}=m\left[a^{2}+c_{1}\left(\frac{a}{m}+c_{1}\right)k^{2}\right]\slashed{v}.
\end{align}
The functions $a$ and $c_{1}$ cannot be constrained further by conditions~\ref{Cond:1} and \ref{Cond:2}.
However we can choose $c_{1}=-a/2m$ to describe non-chiral interactions.
Then, from an off-shell point of view, the function $a$ simply corresponds to the (potentially non-local) field redefinition $Q\rightarrow Q/a$ in the spin-1/2 HPET Lagrangian.
We are free to redefine our fields such that $a=1$. The final result is
\begin{subequations}
\begin{align}
    |\mathbf{p}_{v}\rangle&=|\mathbf{p}\rangle-\frac{\slashed{k}}{2m}|\mathbf{p}]\label{eq:BotUpVarsAngFinal}, \\
    |\mathbf{p}_{v}]&=|\mathbf{p}]-\frac{\slashed{k}}{2m}|\mathbf{p}\rangle\label{eq:BotUpVarsSqFinal}.
\end{align}
\end{subequations}
Thus we recover the on-shell HPET variables in eq.~\eqref{eq:HQETOnShellVars}. 
We conclude that, up to scaling by an overall function of $k^{2}$, \cref{eq:HQETOnShellVars} is the unique
decomposition in terms of traditional variables of non-chiral heavy particle states.
The overall scalings correspond to field redefinitions in the Lagrangian formulation.


\section{Reparameterization and the little group}
\label{sec:RepLG}

As is apparent from eq.~\eqref{eq:HQETMomDecomp}, reparameterization transformations leave $p^{\mu}$ unchanged.
It is therefore reasonable to expect that there exists a relation between reparameterizations and the little group of $p^{\mu}$.
There is indeed a relationship between infinitesimal little group transformations of ${\lambda_{\alpha}}^{I}$ and ${\tilde{\lambda}^{\dot{\alpha}}}_{I}$ and reparameterizations of the total momentum.
The focus of this section is the derivation of such a connection, which is easy
to explore by employing the so-called Light Cone Decomposition (LCD) \cite{KLEISS1985235,Kosower:2004yz} of massive momenta.


The LCD allows any massive momentum to be written as a sum of two massless momenta.
That is, for a momentum $p^{\mu}$ of mass $m$, there exist two massless momenta $a^{\mu}$ and $b^{\mu}$ such that
\begin{align}
    p^{\mu}=a^{\mu}+b^{\mu}.\label{eq:LCD}
\end{align}
When $p^{\mu}$ is real, we can assume wihtout loss of generality that $a^{\mu}$ and $b^{\mu}$ are real as well, since any imaginary components must cancel anyway. The condition $p^{2}=m^{2}$ then implies $a\cdot b=m^{2}/2$. Expressing this in on-shell variables,
\begin{subequations}
\begin{align}
    p_{\alpha\dot{\alpha}}&={\lambda_{\alpha}}^{I}\tilde{\lambda}_{\dot{\alpha}I}=|a\rangle_{\alpha}[a|_{\dot{\alpha}}+|b\rangle_{\alpha}[b|_{\dot{\alpha}}, \\
    p^{\dot{\alpha}\alpha}&={\tilde{\lambda}^{\dot{\alpha}}}_{I}\lambda^{\alpha I}=|a]^{\dot{\alpha}}\langle a|^{\alpha}+|b]^{\dot{\alpha}}\langle b|^{\alpha}.
\end{align}
This allows us to make the identifications
\begin{align}
    {\lambda_{\alpha}}^{1}=|a\rangle_{\alpha},\quad {\lambda_{\alpha}}^{2}=|b\rangle_{\alpha},&\quad \tilde{\lambda}_{\dot{\alpha}1}=[a|_{\dot{\alpha}},\quad\tilde{\lambda}_{\dot{\alpha}2}=[b|_{\dot{\alpha}}.
\end{align}
\end{subequations}

In the spirit of the momentum decomposition in eq.~\eqref{eq:HQETMomDecomp} we can break this up into a large and a small part
\begin{align}
    p^{\mu}&=\alpha a^{\mu}+\beta b^{\mu}+(1-\alpha)a^{\mu}+(1-\beta)b^{\mu},
\end{align}
where $|\alpha|,|\beta|\sim1$. We identify
\begin{align}
    mv^{\mu}\equiv\alpha a^{\mu}+\beta b^{\mu},&\quad k^{\mu}\equiv(1-\alpha)a^{\mu}+(1-\beta)b^{\mu}.
\end{align}
Since $v^{\mu}$ is a four-velocity, it must satify $v^{2}=1$, which constrains $\alpha$ and $\beta$ to obey $\alpha\beta=1$. Once we require this, the on-shell condition that $2mv\cdot k=-k^{2}$ is automatically imposed.


Now, consider a reparameterization of the momentum as in eq.~\eqref{eq:Reparam}. We can use the LCD to rewrite the shift momentum as
\begin{align}
    \delta k^{\mu}&=c^{\mu}+d^{\mu},
\end{align}
where $|c+d|/m\ll1$. For this to be a reparameterization, the new velocity $v^{\mu}+\delta k^{\mu}/m$ must have magnitude $1$, which means $c^{\mu}$ and $d^{\mu}$ must be such that 
\begin{align}\label{eq:RepOnShellness}
    (\alpha a+\beta b)\cdot(c+d)=-c\cdot d.
\end{align}
Contracting the shift momentum with the gamma matrices and using the Schouten identity,
\begin{align}\label{eq:RepMom}
    \delta k_{\alpha\dot{\alpha}}&=\frac{2}{m^{2}}b\cdot(c+d)|a\rangle_{\alpha}[a|_{\dot{\alpha}}+\frac{2}{m^{2}}a\cdot(c+d)|b\rangle_{\alpha}[b|_{\dot{\alpha}}\notag \\
    &\qquad-\frac{[a|(\slashed{c}+\slashed{d})|b\rangle}{m^{2}}|a\rangle_{\alpha}[b|_{\dot{\alpha}}-\frac{[b|(\slashed{c}+\slashed{d})|a\rangle}{m^{2}}|b\rangle_{\alpha}[a|_{\dot{\alpha}}.
\end{align}
Note that setting $k=0$ is always allowed for an on-shell momentum by reparameterization: indeed, choosing $c^{\mu}=(1-\alpha)a^{\mu}$ and $d^{\mu}=(1-\beta)b^{\mu}$ trivially satisfies eq.~\eqref{eq:RepOnShellness}.

Consider an infinitesimal little group transformation of the on-shell variables ${W^{I}}_{J}$ where $W\in SU(2)$. Then we can write
\begin{align}
    {W^{I}}_{J}&={\mathbb{I}^{I}}_{J}+i\epsilon^{j}{U^{jI}}_{J},
\end{align}
where $\epsilon^{j}$ are real and infinitesimal parameters, and ${U^{jI}}_{J}$ is traceless and Hermitian. We suppress the color index $j$ below. Under this transformation, the on-shell variables transform as \cite{Arkani-Hamed:2017jhn}
\begin{subequations}
\begin{align}
    {\lambda_{\alpha}}^{I}&\rightarrow{W^{I}}_{J}{\lambda_{\alpha}}^{J}, \\
    \tilde{\lambda}_{\dot{\alpha}I}&\rightarrow{(W^{-1})^{J}}_{I}\tilde{\lambda}_{\dot{\alpha}J}.
\end{align}
\end{subequations}
Up to linear order in the infinitesimal parameter, the momentum transforms as
\begin{align}
    p_{\alpha\dot{\alpha}}={\lambda_{\alpha}}^{I}\tilde{\lambda}_{\dot{\alpha}I}&\rightarrow(1+i\epsilon{U^{1}}_{1}){\lambda_{\alpha}}^{1}\tilde{\lambda}_{\dot{\alpha}1}+(1+i\epsilon{U^{2}}_{2}){\lambda_{\alpha}}^{2}\tilde{\lambda}_{\dot{\alpha}2}+i\epsilon{U^{2}}_{1}\lambda^{1}_{\alpha}\tilde{\lambda}_{\dot{\alpha}2}+i\epsilon{U^{1}}_{2}\lambda^{2}_{\alpha}\tilde{\lambda}_{\dot{\alpha}1}\notag \\
    &\quad-i\epsilon{U^{2}}_{1}\lambda^{1}_{\alpha}\tilde{\lambda}_{\dot{\alpha}2}-i\epsilon{U^{1}}_{2}\lambda^{2}_{\alpha}\tilde{\lambda}_{\dot{\alpha}1}-i\epsilon{U^{1}}_{1}{\lambda_{\alpha}}^{1}\tilde{\lambda}_{\dot{\alpha}1}-i\epsilon{U^{2}}_{2}{\lambda_{\alpha}}^{2}\tilde{\lambda}_{\dot{\alpha}2}.
\end{align}
Comparing with eq.~\eqref{eq:RepMom}, we would like to identitfy the following map to the reparameterization in eq.~\eqref{eq:Reparam}:
\begin{align}
    i\epsilon {U^{I}}_{J}&\rightarrow{R^{I}}_{J}\equiv\frac{1}{m}\begin{pmatrix}
    2b\cdot\frac{\delta k}{m} & -[b|\frac{\delta k}{m}|a\rangle \\
    -[a|\frac{\delta k}{m}|b\rangle & 2a\cdot\frac{\delta k}{m}
    \end{pmatrix}.
\end{align}
The reparameterization matrix ${R^{I}}_{J}$ is infintesimal because of the appearance of $\delta k^{\mu}/m$ in each entry.
Moreover, ${R^{I}}_{J}$ is traceless up to corrections of order $\mathcal{O}(\delta k^{2}/m^{2})$ because of eq.~\eqref{eq:RepOnShellness}.
However, we cannot equate it to $i\epsilon{U^{I}}_{J}$ because the latter is always anti-Hermitian, whereas ${R^{I}}_{J}$ need not be. 
Indeed, when $\delta k^{\mu}$ is real ${R^{I}}_{J}$ is Hermitian, and when $\delta k^{\mu}$ is imaginary it is anti-Hermitian. 
It can thus be seen that the condition for equality is that $\delta k^{\mu}$ is imaginary:
\begin{align}
    \delta k^{\mu}\in i\mathbb{R}\Rightarrow{\mathbb{I}^{I}}_{J}+{R^{I}}_{J}\in SU(2),
\end{align}
where ${\mathbb{I}^{I}}_{J}+{R^{I}}_{J}$ induces the reparameterization in eq.~\eqref{eq:Reparam}. It is straightforward to check that this quantity also has determinant 1, up to infinitesimal corrections of order $\mathcal{O}(\delta k^{2}/m^{2})$.


\section{Propagators}\label{sec:Propagators}

In ref.~\cite{Durieux:2019eor}, massive on-shell variables were used to construct propagators for massive spin-1/2 and spin-1 states. 
In this section, we use the on-shell HPET variables to do the same for a spin $s\leq2$ state. We find that the propagator for a heavy particle with spin $s\leq2$ is
\begin{align}
    D_{v}^{s}(p_{v})&=P^{s}\frac{N^{s}(p_{v})}{p^{2}-m^{2}}P^{s},
\end{align}
where $P^{s}$ is the spin-$s$ projection operator whose eigenstate is the HPET state, and $N^{s}(p_{v})$ is the numerator of the propagator for a massive particle of that spin.
By recognizing the form of the numerator, this will allow us to extract the higher spin projection operators. The methods used in this section can be applied to arbitrary spin, but become quite cumbersome as the number of little group invariant objects that must be computed grows as $s+1/2$ for half-integer spins, and as $s$ for integer spins. Nevertheless, we are able to use our results to conjecture projection operators for any spin.


\subsubsection*{Spin-1/2}

We begin with the spin-1/2 propagator, which can be constructed as
\begin{align}
	\frac{1}{p^2-m^2} \left[ \begin{pmatrix}
			|p_v^I\rangle \\
		| p_v^I ]
		\end{pmatrix}
		\epsilon_{IJ} 
		\begin{pmatrix}
		\langle -p_v^J | & [- p_v^J |
		\end{pmatrix}
	\right]
=P_{+} \frac{2m_{k}}{p^{2} - m^{2}} P_{+}= P_{+} \frac{1}{\slashed p - m} P_{+}.
\end{align}
%
We do indeed recover the projection operator for a heavy spin-1/2 field.
%


\subsubsection*{Spin-1}

We can do the same for a massive spin-1 field. In this case, we posit that the polarization vector is obtained by replacing $p\rightarrow p_{v}$ and $m\rightarrow m_{k}$ in the usual polarization vector:
\begin{align}
    \varepsilon^{IJ}_{v,\mu}(p)&=\frac{1}{2\sqrt{2}m_{k}}(\langle p^{I}_{v}|\gamma_{\mu}|p^{J}_{v}]+\langle p^{J}_{v}|\gamma_{\mu}|p^{I}_{v}]).
\end{align}
It is straightforward to see that the polarization vector satisfies the requisite condition on the heavy spin-1 particle, $v\cdot\varepsilon^{IJ}_{v}=0$ for $p^{\mu}=mv^{\mu}+k^{\mu}$, as well as the orthonormality condition
\begin{align}
    \varepsilon^{IJ}_{v}\cdot\varepsilon^{LK}_{v}&=-\frac{1}{2}(\epsilon^{IL}\epsilon^{JK}+\epsilon^{IK}\epsilon^{JL}).
\end{align}
The heavy spin-1 propagator is
\begin{align}
	\frac{1}{p^2-m^2}\left[
		\varepsilon^{IJ}_{v,\mu}(p) \epsilon_{IK}\epsilon_{JL} \varepsilon^{LK}_{v,\nu}(-p)
	\right]
	= ( {g_{\mu}}^{\lambda}- v_{\mu} v^{\lambda}) \frac{-g_{\lambda\sigma}+ v_{\lambda}v_{\sigma}}{p^2-m^2}\left({g^{\sigma}}_{\nu} - v^{\sigma}v_{\nu}\right).
\end{align}
From this we can read off that the operator projecting onto the heavy spin-1 particle is $P^{\mu\nu}_{-}$ in \Cref{sec:HPETMatch}.


\subsubsection*{Spin-3/2}

The spin-3/2 polarization tensor is
\begin{align}
    \varepsilon^{IJK}_{v,\mu}(p)=\varepsilon_{v,\mu}^{(IJ}u^{K)}_{v}=\frac{1}{\sqrt{2}m_{k}}\langle p_{v}^{(I}|\gamma_{\mu}|p_{v}^{J}]\begin{pmatrix}
    |p_{v}^{K)}\rangle \\
    |p_{v}^{K)}]
    \end{pmatrix},
\end{align}
where the round brackets around sets of indices denote normalized symmetrization over the indices. Using the symmetry of the spin-1 polarization vector in its little group indices, we have that 
\begin{align}
    \varepsilon^{IJK}_{v,\mu}(p)=\frac{1}{3}\left(\varepsilon^{IJ}_{v,\mu}u_{v}^{K}+\varepsilon^{JK}_{v,\mu}u_{v}^{I}+\varepsilon^{IK}_{v,\mu}u_{v}^{J}\right).
\end{align}
The propagator is
\begin{align}
	\frac{1}{p^2-m^2}&\left[
		\varepsilon^{IJK}_{v,\mu}(p) \epsilon_{IA}\epsilon_{JB}\epsilon_{KC} \varepsilon^{ABC}_{v,\nu}(-p)
	\right]\notag \\
	&= \frac{1}{p^2-m^2}\frac{1}{3}\left(\varepsilon^{IJ}_{v,\mu}\varepsilon_{v,\nu IJ}u^{K}_{v}\bar{u}_{v,K}+2\varepsilon^{IJ}_{v,\mu}\varepsilon_{v,\nu IK}u^{K}_{v}\bar{u}_{v,J}\right)\notag \\
	&=-P_{+}P_{-,\mu\alpha}\frac{2m_{k}}{p^2-m^2}\left[g^{\alpha\beta}-\frac{1}{3}\gamma^{\alpha}\gamma^{\beta}-\frac{1}{3}(\slashed{v}\gamma^{\alpha}v^{\beta}+v^{\alpha}\gamma^{\beta}\slashed{v})\right]P_{-,\beta\nu}P_{+}.
\end{align}
We recognize the quantity between the projection operators as the propagator for a massive spin-3/2 particle with momentum $m_{k}v^{\mu}$ \cite{Behrends:Spin32,Williams:Spin32}. The heavy spin-3/2 projection operator can thus be identified as
\begin{align}
    P^{\mu\nu}_{\frac{1}{2},-}&\equiv P_{+}P_{-}^{\mu\nu}.
\end{align}
%


\subsubsection*{Spin-2}

The spin-2 polarization tensor is
\begin{align}
    \varepsilon^{I_{1}J_{1}I_{2}J_{2}}_{v,\mu_{1}\mu_{2}}(p)=\varepsilon_{v,\mu_{1}}^{(I_{1}J_{1}}\varepsilon^{I_{2}J_{2})}_{v,\mu_{2}}=\frac{1}{2m_{k}^{2}}\langle p_{v}^{(I_{1}}|\gamma_{\mu_{1}}|p_{v}^{J_{1}}]\langle p_{v}^{I_{2}}|\gamma_{\mu_{2}}|p_{v}^{J_{2})}].
\end{align}
Using the symmetry of each spin-1 polarization vector in its little group indices, we find that
\begin{align}
    \varepsilon^{I_{1}J_{1}I_{2}J_{2}}_{v,\mu_{1}\mu_{2}}(p)=\frac{1}{3}\left(\varepsilon_{v,(\mu_{1}}^{I_{1}J_{1}}\varepsilon^{I_{2}J_{2}}_{v,\mu_{2})}+\varepsilon_{v,(\mu_{1}}^{I_{1}I_{2}}\varepsilon^{J_{1}J_{2}}_{v,\mu_{2})}+\varepsilon_{v,(\mu_{1}}^{I_{1}J_{2}}\varepsilon^{I_{2}J_{1}}_{v,\mu_{2})}\right).
\end{align}
The propagator is
\begin{align}
	\frac{1}{p^2-m^2}&\left[
		\varepsilon^{I_{1}J_{1}I_{2}J_{2}}_{v,\mu\nu}(p) \epsilon_{I_{1}K_{1}}\epsilon_{J_{1}L_{1}}\epsilon_{I_{2}K_{2}}\epsilon_{J_{2}L_{2}}\varepsilon^{K_{1}L_{1}K_{2}L_{2}}_{v,\alpha\beta}(-p)
	\right]\notag \\
	&= \frac{1}{p^2-m^2}\frac{1}{3}\left(\varepsilon^{I_{1}J_{1}}_{v,(\mu}\varepsilon^{I_{2}J_{2}}_{v,\nu)}\varepsilon_{v,\alpha I_{1}J_{1}}\varepsilon_{v,\beta I_{2}J_{2}}+2\varepsilon^{I_{1}J_{1}}_{v,(\mu}\varepsilon_{v,\nu)}^{I_{2}J_{2}}\varepsilon_{v,\alpha I_{1}J_{2}}\varepsilon_{v,\beta I_{2}J_{1}}\right)\notag \\
	&=\frac{1}{p^2-m^2}P_{-,\mu\mu^{\prime}}P_{-,\nu\nu^{\prime}}\left[-\frac{1}{2}(P_{-}^{\mu^{\prime}\alpha^{\prime}}P_{-}^{\nu^{\prime}\beta^{\prime}}+P_{-}^{\mu^{\prime}\beta^{\prime}}P_{-}^{\nu^{\prime}\alpha^{\prime}})+\frac{1}{3}P_{-}^{\mu^{\prime}\nu^{\prime}}P_{-}^{\alpha^{\prime}\beta^{\prime}}\right]P_{-,\alpha^{\prime}\alpha}P_{-,\beta^{\prime}\beta}.
\end{align}
The quantity in square brackets is the numerator of the massive spin-2 propagator with momentum $m_{k}v^{\mu}$ \cite{Hinterbichler:2011tt}. We therefore identify the heavy spin-2 projection operator:
\begin{align}
    P^{\mu\nu,\alpha\beta}_{-}&\equiv P^{\mu\nu}_{-}P^{\alpha\beta}_{-}.
\end{align}


\subsection{Spin-\texorpdfstring{$s$}{s} Projection Operator}

Based on the above discussion, as well as the properties of a general spin heavy field, 
we conjecture the projection operator for a spin-$s$ field. An integer spin-$s$ field $Z^{\mu_{1}\dots\mu_{s}}$ must be symmetric and traceless \cite{Singh:1974qz}. 
When the mass of the particle is very large, the particle component $\mathcal{Z}$ must satisfy \cite{Heinonen:2012km}
\begin{align}
	\label{eq:vZ}
    v_{\mu_{1}}\mathcal{Z}^{\mu_{1}\dots\mu_{s}}=0.
\end{align}
By symmetry, this condition holds regardless of the index with which the velocity is contracted. The general spin-$s$ projection operator for a field satisfying \cref{eq:vZ}, and which reduces to the above cases for $s=1$ and $s=2$ is
\begin{align}\label{eq:IntSProj}
    P^{\mu_{1}\nu_{1},\dots,\mu_{s}\nu_{s}}_{-}=\prod_{i=1}^{s}P^{\mu_{i}\nu_{i}}_{-}.
\end{align}
The integer spin projection operator is simply a product of spin-1 projection operators.

A half-integer spin-$(s+1/2)$ field $\Psi^{\mu_{1}\dots\mu_{s}}$ must be symmetric and $\gamma$-traceless \cite{Singh:1974rc},
\begin{align}
    \gamma_{\mu_{1}}\Psi^{\mu_{1}\dots\mu_{s}}=0.
\end{align}
Symmetry ensures that the condition holds for any index the $\gamma$ matrix is contracted with. When the mass of the field becomes very large, its particle component $\mathcal{Q}$ must satisfy \cite{Heinonen:2012km}
\begin{align}
    \slashed{v}\mathcal{Q}^{\mu_{1}\dots\mu_{s}}=\mathcal{Q}^{\mu_{1}\dots\mu_{s}}.
\end{align}
These constraints also imply, among other things, the $v$-tracelessness of the heavy field. The general spin-$(s+1/2)$ projection operator that results in a field satisfying these conditions, and that reduces to the above cases for spin-1/2 and spin-3/2, is
\begin{align}\label{eq:HalfIntSProj}
    P^{\mu_{1}\nu_{1},\dots,\mu_{s}\nu_{s}}_{\frac{1}{2},-}\equiv P_{+}P^{\mu_{1}\nu_{1},\dots,\mu_{s}\nu_{s}}_{-}.
\end{align}
From this we see that knowledge of the spin-1/2 heavy particle states
is enough to construct the polarization tensors and projection operators for higher spin states.
In this sense, HPETs are unified in terms of the basic building blocks in \cref{eq:HQETOnShellVars}.


\section{Matching to HPET Lagrangians}\label{sec:HPETMatch}

In this section, we address the matching of on-shell
amplitudes to those derived from HPET Lagrangians.
First, there is a subtlety that must be accounted for when matching the minimal coupling in eqs.~\eqref{eq:OnShellDictAllk}
and \eqref{eq:OnShellDictAllkSq} to an HPET Lagrangian. We focus the discussion of this to the case of spin-1/2 HPET.
Next, we confirm explicitly that the general spin
three-point amplitude derived from the Zeeman coupling in ref.~\cite{Chung:2018kqs}
reproduces the amplitude derived from spin-1 abelian HQET
when expressed using on-shell HPET variables.

\subsection{Matching spin-1/2 minimal coupling}

For any quantum field theory, the form of the Lagrangian that produces a given $S$-matrix is not unique: indeed the $S$-matrix is invariant under appropriate redefinitions of the fields composing the Lagrangian \cite{Manohar:2018aog}. Generally, a field redefinition will alter the Green's function for a given process. To relate the Green's functions of two forms of a Lagrangian, the relation between both sets of external states must be specified. The same holds for HQET, which has been presented in various forms in the literature.

Fortunately, the definition of the heavy spinors in eq.~\eqref{eq:HQETSpinor} specifies for us the form of the spin-1/2 HPET Lagrangian whose external spinors are expressible as such. By inverting eq.~\eqref{eq:HQETSpinor}, we see that 
the field redefinition converting the full theory to its HPET form must reduce to \begin{align}\label{eq:PosSpaceDef}
    \psi(x)&=e^{-imv\cdot x}\left[\frac{1+\slashed{v}}{2}+\frac{1-\slashed{v}}{2}\frac{1}{iv\cdot\partial+2m}i\slashed{\partial}\right]Q_{v}(x).
\end{align} 
in the free-field limit. 
For spin-1/2 HQET, this means we must match the minimal coupling to the Lagrangian in the form
\begin{align}
    \mathcal{L}^{s=\frac{1}{2}}_{\text{HQET}}=\bar{Q}iv\cdot DQ+\bar{Q}i\slashed{D}P_{-}\frac{1}{2m+iv\cdot D}i\slashed{D}Q.
\end{align}
This form of the Lagrangian appears in e.g. ref.~\cite{CHEN1993421}, and differs from the forms in refs.~\cite{Damgaard:2019lfh,Manohar:1997qy} by the presence of a projection operator in the non-local term. The Lagrangian of HBET presented in ref.~\cite{Damgaard:2019lfh} must similarly be modified to compare to the minimal coupling amplitude. The suitable form for spin-1/2 HBET is
\begin{subequations}\label{eq:HBETLagLocal}
\begin{align}
    \mathcal{L}_{\textrm{HBET}}^{s=\frac{1}{2}}&=\sqrt{-g}\bar{Q}i\mathcal{D}Q+\frac{\sqrt{-g}}{2m}\bar{Q}i\mathcal{D}P_{-}
    \sum_{n=0}^{\infty} G_n[h]\frac{F[h]^n}{m^n}
    i\mathcal{D}Q,
\end{align}
where
\begin{align}
    i\mathcal{D}&\equiv i{e^{\mu}}_{a}\gamma^{a}\,D_{\mu} 
    +mv_{\mu}\gamma^{a}({e^{\mu}}_{a}-\delta^{\mu}_{a}),
\end{align}
and all other notation is described in ref.~\cite{Damgaard:2019lfh}.
\end{subequations}

\subsection{Matching spin-1 Zeeman coupling}

We demonstrate explicitly the applicability of the on-shell HPET variables
to spin-1 heavy particle systems.
To do so, we will show that the same variables are suitable for describing
the three-point amplitude arising from the Proca action.
First, we note that a massive spin-1 particle described by the Proca action
has a gyromagnetic ratio $g=1$ \cite{Holstein:2006wi}.
As such, it should not be expected that the corresponding three-point
amplitude matches with the minimal coupling amplitude for $s=1$.
To understand which three-point amplitude we should match with,
we recast the three-point amplitude derived from the Zeeman coupling
in ref.~\cite{Chung:2018kqs} into on-shell HPET variables (with $k_{1}=0$):
\begin{align}
    \cA^{+,s}&=\frac{g_{0}x}{m^{2s}}\left[\langle\mathbf{2}_{v}\mathbf{1}_{v}\rangle^{2s}+x\frac{sg}{2m}\langle\mathbf{2}_{v}\mathbf{1}_{v}\rangle^{2s-1}\langle\mathbf{2}_{v}3\rangle\langle3\mathbf{1}_{v}\rangle+\dots\right],
\end{align}
where the dots represent higher spin multipoles. 
When $g=2$ we recover the spin-dipole term from $2s$ factors of the spin-1/2 minimal coupling amplitude.
Setting $s=g=1$ for the Proca action,
\begin{align}\label{eq:ZeemanProca}
    \cA^{+,1}&=\frac{g_{0}x}{m^{2}}\left[\langle\mathbf{2}_{v}\mathbf{1}_{v}\rangle^{2}+\frac{x}{2m}\langle\mathbf{2}_{v}\mathbf{1}_{v}\rangle\langle\mathbf{2}_{v}3\rangle\langle3\mathbf{1}_{v}\rangle+\dots\right].
\end{align}
This is the three-point amplitude that we expect from a very heavy spin-1 Proca particle.

Consider now the Proca Lagrangian for a massive vector field $B^{\mu}$ coupled to electromagnetism:
\begin{subequations}
\begin{align}
    \mathcal{L}&=-\frac{1}{4}F^{*}_{\mu\nu}F^{\mu\nu}+\frac{1}{2}m^{2}B^{*}_{\mu}B^{\mu},
\end{align}
where
\begin{align}
F^{\mu\nu}=D^{\mu}B^{\nu}-D^{\nu}B^{\mu},\quad D^{\mu}B^{\nu}=(\partial^{\mu}+ieA^{\mu})B^{\nu},
\end{align}
\end{subequations}
and $A^{\mu}$ is the $U(1)$ gauge field.
We now need a condition that splits the light component $\mathcal{B}^{\mu}$ from the heavy (anti-field) component $\tilde{\mathcal{B}}^{\mu}$. Furthermore, the light component has to satisfy $v_{\mu}\mathcal{B}^{\mu}=0$ \cite{Heinonen:2012km}. The appropriate decomposition of the massive vector field is
\begin{subequations}\label{eq:HeavySpin1}
\begin{align}
    \mathcal{B}^{\mu}&=e^{imv\cdot x}P^{\mu\nu}_{-}B_{\nu}, \\
    \tilde{\mathcal{B}}^{\mu}&=e^{imv\cdot x}P^{\mu\nu}_{+}B_{\nu},
\end{align}
\end{subequations}
where $P^{\mu\nu}_{-}\equiv g^{\mu\nu}- v^{\mu}v^{\nu}$ --- this is the projection operator that has been derived explicitly in Appendix~\ref{sec:Propagators} ---
and $P^{\mu\nu}_{+}\equiv v^{\mu}v^{\nu}$.
Next, we substitute \cref{eq:HeavySpin1} into the Proca Lagrangian,
and integrate out $\tilde{\mathcal{B}}^{\mu}$ using its equation of motion to find
\begin{align}
	\mathcal{L}^{s=1}_{\text{HQET}}&= -m \mathcal{B}_\mu^* (iv\cdot D) \mathcal{B}^\mu 
	- \frac{1}{4} \mathcal{B}_{\mu\nu}^* \mathcal{B}^{\mu\nu} 
	+ \frac{1}{2} \mathcal{B}_\nu^* D^\nu D_\mu \mathcal{B}^\mu 
	+ \mathcal{O}(m^{-1}),
\end{align}
where $\mathcal{B}^{\mu\nu}=D^\mu\mathcal{B}^\nu - D^\nu\mathcal{B}^\mu$.
Computing the three-point amplitude with this Lagrangian for $k_{1}=0$ and expressing
it using on-shell HPET variables, we find agreement with \cref{eq:ZeemanProca} for $g_{0}=-em/\sqrt{2}$.
This supports the hypothesis that the on-shell information of spin-1/2 HPET is sufficient to extend HPETs to higher spins.

\bibliographystyle{JHEP}
\bibliography{OSHQET}

\end{document}